\documentclass{aa}  

\usepackage{algorithm,algorithmic}
\usepackage[T1]{fontenc}
\usepackage{ae,aecompl}
\usepackage{natbib}

\usepackage{graphicx,epsfig,amsmath,color}
\usepackage{txfonts}
\newcommand{\beq}{\begin{equation}}
\newcommand{\eeq}{\end{equation}}
\newcommand{\beqn}{\begin{eqnarray}}
\newcommand{\eeqn}{\end{eqnarray}}

\def\bmath#1{\mbox{\boldmath$#1$}}

\long\def\symbolfootnote[#1]#2{\begingroup%
\def\thefootnote{\fnsymbol{footnote}}\footnote[#1]{#2}\endgroup}

\usepackage{pstricks,pst-text}
\floatname{algorithm}{Algorithm}

\begin{document} 

   \title{Diffuse radio sky models using large scale shapelets}
   \titlerunning{Diffuse radio sky models using shapelets}
     
   \author{S.~Yatawatta\inst{1},
          }

\institute{
      ASTRON, Netherlands Institute for Radio Astronomy, Oude Hoogeveensedijk 4, 7991 PD, Dwingeloo, The Netherlands.
    }
    \authorrunning{S.~Yatawatta}
   \date{Received ---; accepted ---}

 
  \abstract
  {}
  {
    Sky models used in radio interferometric data processing primarily consist of compact and discrete radio sources. When there is a need to model large scale diffuse structure such as the Galaxy, specialized source models are sought after for the sake of simplicity and computational efficiency. In this paper, we propose the use of shapelet basis functions for modeling large scale diffuse structure in various radio interferometric data processing pipelines.
  }
  {
    Conventional source model construction using shapelet basis functions is restricted to using images of smaller size due to limitations in computational resources such as memory. We propose a novel shapelet decomposition method to lift this restriction, enabling the use of images of millions of pixels (as well as a wide spectral bandwidth) for building models of large scale diffuse structure. Furthermore, the application of direction dependent errors onto diffuse sky models is an expensive operation often performed as a convolution. In this paper, we propose using some specific properties of shapelet basis functions to apply such direction dependent errors as a product of the model coefficients, thus the avoiding the need for convolution.
  }
  {
    We provide results based on simulations and real observations. In order to measure the efficacy of our proposed method in modeling large scale diffuse structure, we consider direction dependent calibration of simulated as well as real LOFAR observations that have a significant amount of large scale diffuse structure. Results show that by including large scale shapelet models of the diffuse sky, we are able to overcome a major problem faced by existing calibration techniques that do not model such large scale diffuse structure, i.e., the suppression of such large scale diffuse structure due to model incompleteness.
  }
  {}

   \maketitle

\section{Introduction\label{intro}}
Apart from compact (extragalactic) radio sources, the low frequency radio sky is also covered by large scale diffuse structure, the most obvious being the Galaxy \citep{Zheng2017diffuse,Byrne2021}. The effect of such large scale diffuse structure on high precision data processing (such as for the study of the Epoch of Reionization (EoR) and the Cosmic Dawn (CD) \citep{SKA1,Bharat2018,Bharat2020}) is considered to be an issue well worthy of thorough investigation \citep{Gehlot2021,Jackson2023,Barry2023}. Such investigations recommend the inclusion of large scale diffuse structure in the sky models used in low frequency interferometric data processing. However, the commonly used data models (for example in radio interferometric calibration) only accommodate compact and discrete radio sources and the inclusion of large scale structure would incur significant computational cost \citep{Carozzi2009}. Therefore, the commonly used calibration strategy is to indirectly minimize the effect of large scale diffuse structure on the performance of calibration. For example, short baselines that are mostly affected by large scale diffuse structure can be excluded while calibration is being performed \citep{Patil2016,Patil2017}. On the other hand, some form of filtering can be pre-applied to the data to remove the signal due to large scale diffuse structure prior to calibration \citep{Charles2023}. Such indirect remedies are not always feasible, for instance when there are not enough long baselines to base the calibration upon \citep{Munshi2023} because excluding short baselines also excludes many usable constraints \citep{EW2017}.

In this paper, we propose a method of directly modeling the large scale diffuse structure in low frequency radio interferometric data using shapelet basis functions \citep{SHP1,SHP5}. The novelty of the proposed work can be summarized as follows: 
\begin{itemize}
  \item Heretofore, shapelet model construction have only used (relatively) small images as input \citep{Shap1,Shap,Line2020PASA}. Modern radio interferometers are capable of covering large fields of view, yielding images of sizes leading up to millions or billions of pixels \citep{Excon}. Using such large scale images as input to shapelet model creation is computationally restricted due to the large memory requirement. In this paper, we overcome this limitation by using a divide and conquer approach by using the accelerated projection based consensus (APC) algorithm \citep{azizan2019distributed}. This not only enables us to work with images of millions of pixels in size (covering a large field of view at high resolution), but also images made at a large number of frequencies covering a wide bandwidth. In other words, we are able to construct both wide-field as well as wide-band shapelet models using a distributed, divide and conquer approach in a computationally feasible manner.
  \item  Interferomety with large fields of view --which is the norm for low frequency interferometers-- inevitably has to encounter systematic errors (such as the beam shape) covering a large field of view. The application of such systematic errors onto diffuse sky models requires convolution in the Fourier space, and that is a computationally expensive operation \citep{Lanman_2022}. In this paper, we overcome the use of convolution by exploiting certain properties of the shapelet basis functions \citep{SHP1,SHP5}, in particular their orthogonality. By exploiting these properties we can perform coordinate-free convolutions entirely using coefficient multiplications of the basis functions, significantly reducing the computational cost.
  \item As an application of large scale shapelet models in radio interferometry, we extend  our previous work on distributed, direction dependent calibration with spectral and spatial regularization \citep{DCAL,spatial_cal} to incorporate models of large scale diffuse sky.
\end{itemize}

In calibration, a commonly used alternative to modeling large scale diffuse structure in the aforementioned manner is to represent such structure by sources with compact support, such as point sources or Gaussians. While this representation scheme is much simpler, it has several drawbacks: (i) A large number of components are needed for accurate representation of large scale diffuse structure (increased computational cost), (ii) The inherent spatial continuity in the diffuse structure is not always guaranteed (less realistic), and (iii) Such models are tuned to one specific observation and their re-use in other observations is not straight forward, for example due to coordinate transformation, projection and uv-coverage (limited re-usability). On the other hand, in image synthesis and deconvolution, a plethora of alternative methods exist that specialize on modeling diffuse structure, e.g.,  \citep{cornwell2008multiscale,Bhatnagar2004,Rich2008,Rau2011,Junk2016} and integration of such methods into direction dependent calibration is worthy of further investigation.

The rest of the paper is organized as follows: In section \ref{sec:model}, we introduce shapelet basis functions and the data model used in radio interferometry. We also introduce the application of direction dependent effects onto shapelet source models as a multiplication of shapelet coefficients (in the image space, instead of a convolution in Fourier space). In section \ref{sec:apc}, we provide details of wide-field and wide-band shapelet model construction using the APC algorithm. In section \ref{sec:calib}, we provide details of the use of large scale diffuse sky models in distributed, direction dependent calibration. We provide results based on simulations and real data in sections \ref{sec:results} and \ref{sec:observations} respectively, to illustrate the use of large scale shapelet models in direction dependent calibration. Finally, we draw our conclusions in section \ref{sec:conclusions}.

{\em Notation}: Lower case bold letters refer to column vectors (e.g., ${\bf y}$). Upper case bold letters refer to matrices (e.g., ${\bf C}$). Unless otherwise stated, all parameters are complex numbers. The set of integers is given as ${\mathbb Z}$, the set of complex numbers is given as ${\mathbb C}$, and the set of real numbers as  ${\mathbb R}$. The matrix inverse, pseudo-inverse, transpose, and Hermitian transpose are referred to as $(\cdot)^{-1}$, $(\cdot)^{\dagger}$, $(\cdot)^{T}$, $(\cdot)^{H}$, respectively. The matrix Kronecker product is given by $\otimes$. The identity matrix of size $N$ is given by ${\bf I}_N$. The Frobenius norm is given by $\|\cdot \|$ and the L-$1$ norm is given by $\|\cdot\|_1$.

\section{Data model\label{sec:model}}
In this section, we introduce the (rectangular) shapelet basis functions, the commonly used radio interferometric data model and how we expand this model to accommodate large scale diffuse structure using shapelet basis functions.

\subsection{Shapelet basis functions\label{sec:shap}}
In this paper, we use the rectangular shapelet basis functions \citep{SHP1,SHP2,SHP3,SHP4} as the basis for sky model construction. One dimensional shapelets or Gauss-Hermite polynomials can be given as
\beq \label{shap}
\phi_n(x,\beta)=\left[ 2^n \sqrt{\pi} n! \beta\right]^{-\frac{1}{2}} H_{n}\left(\frac{x}{\beta}\right) \exp\left(-0.5 \left(\frac{x}{\beta}\right)^2\right)
\eeq
where $x$ ($\in \mathbb{R}$) is the coordinate, $n$ ($\mathbb{Z}^{+}$) is the order of the basis function and  $\beta$ ($\in\mathbb{R}^{+}$) is the scale factor. The Hermite polynomial of order $n$ is given by $H_n(\cdot)$. In two dimensions (with coordinates $x$ and $y$), the basis is constructed as a product of one dimensional basis functions, i.e., $\phi_n(x,y,\beta)=\phi_n(x,\beta)\phi_n(y,\beta)$.

Shapelet basis functions are already being used for radio astronomical source modeling in practice (see e.g., \cite{Shap1,Shap,Line2020PASA}). One noteworthy property that we exploit is the fact that the Fourier transform of $\phi_n(x,\beta)$ is $\jmath^{n} \phi_n(u,1/\beta)$ where $u$ is the Fourier dual of $x$. In other words, given the shapelet decomposition in real (image) space, we can calculate its Fourier transform with linear computational complexity.

We state one additional property of rectangular shapelet basis functions that will be useful later.
\begin{lemma}
  Given functions $f(x)$ and $g(x)$ and their shapelet decompositions, $f(x)$: scale $\beta_f$, coefficients $f_n$, $n=0,\ldots,M_f-1$, $g(x)$: scale $\beta_g$, coefficients $g_m$, $m=0,\ldots,M_g-1$, the shapelet decomposition of the product $h(x)=f(x)g(x)$ with scale $\beta_h$ and coefficients $h_l$, $l=0,\ldots,M_h-1$ can be given in closed form by (\ref{shap_product}). 
\end{lemma}
The coefficients of the product are
\beq \label{shap_product}
h_l=\sum_{m=0}^{M_g-1}\sum_{n=0}^{M_f-1} C_{lmn}(\beta_h,\beta_f,\beta_g) f_n g_m
\eeq
where $C_{lmn}(\beta_h,\beta_f,\beta_g)$ is given by \citep{SHP5}
\beqn \label{Clmn}
\lefteqn{
  C_{lmn}(a,b,c)=}\\\nonumber
&&\nu \left[ (-2)^{l+m+n} \sqrt{\pi} l! m! n! a b c \right]^{-\frac{1}{2}}  L_{l,m,n}\left(\sqrt{2}\frac{\nu}{a},\sqrt{2}\frac{\nu}{b},\sqrt{2}\frac{\nu}{c}\right)
\eeqn
  for even values of $l+m+n$ and $\nu=\frac{1}{\sqrt{a^{-2}+b^{-2}+c^{-2}}}$, $a,b,c\in \mathbb{R}^{+}$.
  Furthermore, $L_{l,m,n}(a,b,c)$ in (\ref{Clmn}) can be evaluated using $L_{0,0,0}(a,b,c)=1$ and the recurrence relations for even values of $l+m+n$,
\beqn \label{Llmn}
  \lefteqn{
    L_{l+1,m,n}(a,b,c)= 2l(a^2-1)L_{l-1,m,n}(a,b,c)}\\\nonumber
  && +2mabL_{l,m-1,n}(a,b,c)+2ncaL_{l,m,n-1}(a,b,c)\\\nonumber
  \lefteqn{
    L_{l,m+1,n}(a,b,c)= 2m(b^2-1)L_{l,m-1,n}(a,b,c)}\\\nonumber
  && +2nbcL_{l,m,n-1}(a,b,c)+2labL_{l-1,m,n}(a,b,c)\\\nonumber
  \lefteqn{
    L_{l,m,n+1}(a,b,c)= 2n(c^2-1)L_{l,m,n-1}(a,b,c)}\\\nonumber
  && +2lcaL_{l-1,m,n}(a,b,c)+2mbcL_{l,m-1,n}(a,b,c)
\eeqn
and with $L_{l,m,n}(a,b,c)=0$ for odd values of $l+m+n$. 

The proof of Lemma 1 can be found by following the proof given in \citep{SHP6}. If we are free to select the scale $\beta_h$ and the number of the basis functions $M_h$ of the product, we can set $\beta_h$ to the lower value of $\beta_f$ and $\beta_g$ and $M_h$ to the higher value of $M_f$ and $M_g$.

The Fourier relationship and the results in Lemma 1 directly extend to the two dimensional basis that will be used for modeling large scale diffuse structure.

\subsection{Radio interferometric data model}
We consider an interferometric array with $N$ receivers. The output visibilities of the baseline formed by stations $p$ and $q$ can be given as \citep{HBS}
\beq \label{V}
{\bf V}_{pqf_i}= \sum_{k=1}^K {\bf J}_{pkf_i}{\bf C}_{pqkf_i}{\bf J}_{qkf_i}^H + {\bf J}_{p0f_i} \mathcal{\bf C}_{pqf_i}({\bf S},{\bf Z}) {\bf J}_{q0f_i}^H + {\bf N}_{pq}
\eeq
where ${\bf V}_{pqf_i}$ ($\in \mathbb{C}^{2\times 2}$) are the observed data at frequency $f_i$, at $P$ distinct frequencies, $i\in[1,P]$. The right hand side of (\ref{V}) describes the model used in predicting the visibilities given the sky model. We consider $K$ distinct directions in the sky, each direction being modeled by a set of compact sources and the coherencies for the $k$-th direction are given by ${\bf C}_{pqkf_i}$ ($\in \mathbb{C}^{2\times 2}$). The systematic errors determined by calibration along the $k$-th direction are given by ${\bf J}_{pkf_i}$ and ${\bf J}_{qkf_i}$ ($\in \mathbb{C}^{2\times 2}$).

The coherencies representing the diffuse sky model are given by the mapping $\mathcal{\bf C}_{pqf_i}({\bf S},{\bf Z})$ ($\{{\bf S},{\bf Z}\} \rightarrow \mathbb{C}^{2\times 2}$) in (\ref{V}). The systematic errors ${\bf J}_{p0f_i}$ and ${\bf J}_{q0f_i}$ ($\in \mathbb{C}^{2\times 2}$) that are attributed to the diffuse sky model are used as an antenna based 'scaling' of the diffuse sky model (hence they are independent of the direction) and the true direction dependence is represented by $\mathcal{\bf C}_{pqf_i}({\bf S},{\bf Z})$. Therefore, the physical relevance of ${\bf J}_{p0f_i}$ and ${\bf J}_{q0f_i}$ is to account for any discrepancy of the total flux in the diffuse sky model and the true sky, mainly because we do not have zero length baselines (or baselines close to zero). We consider $\mathcal{\bf C}_{pqf_i}({\bf S},{\bf Z})$ to be a function of two tensors, namely ${\bf S}$ and ${\bf Z}$, with ${\bf S}$ describing the source model and ${\bf Z}$ describing the spatial model (such as the beam shape and ionosphere). One noteworthy property that is common to both ${\bf S}$ and to ${\bf Z}$ is the use of shapelet basis functions whose coefficients are given by ${\bf S}$ and ${\bf Z}$ while evaluating $\mathcal{\bf C}_{pqf_i}({\bf S},{\bf Z})$. The origin of the diffuse sky model ($x=y=0$) can in general be anywhere in the sky (this position is used to calculate the phase contribution), but for simplicity, we consider its origin to coincide with the phase center of the observation. Moreover, both ${\bf S}$ and ${\bf Z}$ can consist of multiple bases (for example with different scales $\beta$ and number of basis functions $M$).

Finally, the noise and the effect of unmodeled sources is given by the last term in (\ref{V}), i.e., ${\bf N}_{pq}$ ($\in \mathbb{C}^{2\times 2}$) which is generally considered to be consisting of elements drawn from a complex, circular Gaussian distribution. Given the data model (\ref{V}), the objective of calibration is to determine ${\bf J}_{pkf_i}$ for all $p$ and $k=0,\ldots,K$. We have formulated calibration as a distributed optimization problem in our previous work \citep{DCAL,spatial_cal} which we briefly describe here. For brevity, we group all solutions for one direction $k$ and one frequency $f_i$ as 
\beq \label{Jf}
{{\bf J}_{kf_i}}\buildrel\triangle\over=[{\bf{J}}_{1k{f_i}}^T,{\bf{J}}_{2k{f_i}}^T,\ldots,{\bf{J}}_{Nk{f_i}}^T]^T,
\eeq
where ${{\bf J}_{kf_i}}$ ($\in \mathbb{C}^{2N\times 2}$) groups the $N$ solutions for all stations into one block.

The core principle in spectrally and spatially regularized direction dependent calibration can be given as 
\beq \label{3fac}
{\bf J}_{kf_i}={\bf B}_{f_i} {\bf Z} {\bmath \Phi}_k
\eeq
where ${\bf B}_{f_i}$ ($\in \mathbb{C}^{2N\times 2FN}$) is a basis in frequency and ${\bmath \Phi}_k$ ($\in \mathbb{C}^{2G \times 2}$) is a basis in space while ${\bf Z}$ ($\in \mathbb{C}^{2FN\times 2G}$) is the 'global' model that can be used to derive all solutions ${\bf J}_{kf_i}$ for all $k$ and $f_i$ \citep{spatial_cal}. The bases can be constructed from any polynomial, for example,
\beq \label{bases}
{\bf B}_{f_i}={\bf b}_{f_i}^T \otimes {\bf I}_{2N},\ \ {\bmath \Phi}_k={\bf I}_2 \otimes {\bmath \phi}_k
\eeq
creates a basis in frequency using ${\bf b}_{f_i}$ ($\in \mathbb{C}^{F\times 1}$) which is a set of $F$ polynomials evaluated at frequency $f_i$ and a basis in space using ${\bmath \phi}_k$ ($\in \mathbb{C}^{G\times 1}$) which is a set of $G$ polynomials in space (2 dimensional) that are evaluated at the coordinates pointing towards the $k$-th direction.

The spectral regularization is achieved by enforcing the constraint 
\beq \label{BZ}
{\bf J}_{kf_i}={\bf B}_{f_i} {\bf Z}_{k}
\eeq
and  the spatial regularization is achieved by enforcing the constraint
\beq \label{Zbar}
 {\bf Z}_k=\overline{{\bf Z}}_k \buildrel \triangle \over = {\bf Z} {\bmath \Phi}_k
\eeq
and we describe the constrained calibration in detail in section \ref{sec:calib}.

The spatial basis functions ${\bmath \phi}_k$ are created using the shapelet basis as described in section \ref{sec:shap}. In order to evaluate $\mathcal{\bf C}_{pqf_i}({\bf S},{\bf Z})$ for any given baseline $p,q$ and frequency $f_i$, we evaluate the shapelet coefficients (using the appropriate scale factors) for the source model given by coefficients ${\bf S}$ and the shapelet coefficients for the spatial model given by  ${\bf Z}$ and select the coefficients for stations $p$ and $q$. Note that (\ref{shap_product}) describes a scalar product, but the same relationship holds when we find a product of coherencies ($\in \mathbb{C}^{2\times 2}$). Note also that we need to apply (\ref{shap_product}) twice, first to find the product of the (Hermitian)  spatial model for station $q$ with the source coherency and thereafter, to find the product of the spatial model for station $p$ with the result of the previous step. Finally, we use the Fourier equivalence of the shapelet basis functions to calculate the visibilities for $\mathcal{\bf C}_{pqf_i}({\bf S},{\bf Z})$.

Looking back at (\ref{V}), we see that the unknowns are not only ${\bf J}_{kf_i}$ given by (\ref{Jf}) but also ${\bf Z}$ which in turn is dependent on ${\bf J}_{kf_i}$. In other words, there is a hidden cyclic dependency and this might lead to instability in calibration. We break this dependency by introducing the constraint
\beq \label{Ztildecons}
\widetilde{\bf Z}={\bf Z}
\eeq
and use $\mathcal{\bf C}_{pqf_i}({\bf S},\widetilde{\bf Z})$ in (\ref{V}) where the only unknowns are ${\bf J}_{kf_i}$. The spatial model used for predicting the diffuse sky model, i.e.,  $\widetilde{\bf Z}$ ($\in \mathbb{C}^{2FN\times 2G}$) in (\ref{Ztildecons}) is updated with a delay as we describe in section \ref{sec:calib}.

Ideally, when there are no direction dependent systematic errors, the spatial model should be isotropic, or in other words, the gain should be unity along all directions. We find the isotropic spatial model $\widetilde{\bf Z}_0$ ($\in \mathbb{C}^{2FN\times 2G}$) as
\beqn \label{Z0}
\lefteqn{
  \widetilde{\bf Z}_0=\underset{{\bf Z}}{\rm arg\ min} \sum_{k,f_i} \| {\bf B}_{f_i} {\bf Z}  {\bmath \Phi}_k - \widetilde{\bf J}\|^2} \\\nonumber
&&=\left( \sum_{f_i} {\bf B}_{f_i}^H {\bf B}_{f_i} \right)^{\dagger} \left( \sum_{f_i} {\bf B}_{f_i}^{H} \right) \widetilde{\bf J} \left(\sum_{k} {\bmath \Phi}_k^H \right) \left( \sum_{k} {\bmath \Phi}_k {\bmath \Phi}_k^H \right)^{\dagger}
\eeqn
where $\widetilde{\bf J}={\bf 1}_N \otimes {\bf I}_2$ with ${\bf 1}_N$ ($\in \mathbb{R}^{N\times 1}$) being a vector of all ones. As we describe in section \ref{sec:calib}, we keep the spatial model used for predicting the diffuse sky model $\widetilde{\bf Z}$ close to $\widetilde{\bf Z}_0$ as much as possible to improve the stability of calibration. This is mostly due to the fact that ${\bf Z}$ can be biased in the case of imperfect sky models (for directions $k=1\ldots,K$) especially when we start with a small number of directions $K$. As we progress, we can increase $K$ to cover the full sky, but initially this is not always possible.

\section{Accelerated projection based consensus\label{sec:apc}}
In this section, we describe shapelet model construction using large images, typically over millions of pixels, as input. For the simplest case, we consider an image of size $N_x \times N_y=N_I$ pixels, that is represented by a vector ${\bf b}\in \mathbb{R}^{N_I\times 1}$. Based on this image, we need to construct a model with $M$ basis functions whose coefficients are given by ${\bf x}\in \mathbb{R}^{M\times 1}$. The basis matrix is given by ${\bf A}\in \mathbb{R}^{N_I\times M}$  where each row of ${\bf A}$ is created by the $M$ basis functions evaluated at the pixel coordinates representing the corresponding row of ${\bf b}$.  The solution for ${\bf x}$ is found by solving  
\beq \label{axb}
{\bf A}{\bf x}={\bf b}
\eeq
which is directly solvable when the size of ${\bf A}$ in (\ref{axb}) is small. However, we consider the situation where ${\bf A}$ cannot fit in memory (let alone finding the inverse of ${\bf A}$) because either $N_I$ or $M$ or both are large. Due to this, we cannot directly solve (\ref{axb}) as in the case for building models for compact diffuse sources. 

In a more practical situation, we create a family of models (with varying values for scale $\beta$ and the number of basis functions $M$) for any given image. In this case, say with $d$ bases, the number of columns in ${\bf A}$ is increased even further (sum of all $M$ for $d$ bases). Furthermore, we can build a model with frequency dependence, where ${\bf b}$ in (\ref{axb}) is a concatenation of images at different frequencies. In this case, the number of rows of ${\bf A}$ is increased as well (the bases at each frequency are evaluated with a known frequency dependence).

In aforementioned situations, instead of directly solving (\ref{axb}), we use a divide and conquer approach. We divide the original problem into $B$ sub-problems as (\ref{axb_div})

\beq \label{axb_div}
\left[ \begin{array}{c}
  {\bf A}_1\\
  {\bf A}_2\\
  \vdots\\
  {\bf A}_B
\end{array} \right]
{\bf x}=
\left[ \begin{array}{c}
{\bf b}_1\\
{\bf b}_2\\
\vdots \\
{\bf b}_B\\
\end{array} \right]
\eeq
where we have divided the original matrix ${\bf A}$  (size $N_I \times M$) into sub-matrices ${\bf A}_i$ (size $n_i\times M$, $n_i\approx N_I/B$) and the original vector ${\bf b}$ (size $N_I$) into sub-vectors ${\bf b}_i$ (size $n_i$), $i\in[1,B]$. Note that the summation $\sum_i n_i=N_I$. 
We apply the accelerated projection based consensus (APC) algorithm \citep{azizan2019distributed} to solve (\ref{axb_div}). The pseudocode for APC is given in Algorithm \ref{algAPC}.
\begin{algorithm}
  \caption{Shapelet model construction using accelerated projection based consensus}
\label{algAPC}
\begin{algorithmic}[1]
  \REQUIRE Number of subtasks $B$, momentum factors $\gamma,\eta\in (0, 1]$, maximum iterations $T$.
  \STATE Divide ${\bf A}$ into $B$ sub-matrices and ${\bf b}$ into $B$ sub-vectors by selecting subsets of rows to get ${\bf A}_i$, ${\bf b}_i$, $i =1\ldots,B$.
  \STATE Find initial solutions, ${\bf x}_i^{(0)}={\bf A}_i^{\dagger}{\bf b}_i$, $\forall i$.
  \STATE Find projection matrices ${\bf P}_i={\bf I}_M-{\bf A}_i^T({\bf A}_i{\bf A}_i^T)^{\dagger}{\bf A}_i$, $\forall i$.
  \STATE Initialize the global solution $\overline{\bf x}^{(0)}\leftarrow{\bf 0}$.
\FOR{$t=0,\ldots,T-1$}
  \STATE Compute agents update in parallel ${\bf x}_i^{(t+1)}\leftarrow {\bf x}_i^{(t)} +\gamma {\bf P}_i\left(\overline{\bf x}^{(t)}-{\bf x}_i^{(t)}\right)$, $\forall i$.
  \STATE Fusion center updates $\overline{\bf x}^{(t+1)}\leftarrow \frac{\eta}{B}\sum_{l=1}^{B} {\bf x}_l^{t+1} + (1-\eta) \overline{\bf x}^{(t)}$.
\ENDFOR
\STATE Return $\overline{\bf x}^{(t+1)}$ as the solution.
\end{algorithmic}
\end{algorithm}

By making $B$ large enough, we can break up any large problem such as (\ref{axb}) into manageable sub-problems. Notice that the $B$ subtasks in Algorithm \ref{algAPC} can be executed in parallel, making this also suitable for distributed implementation. The most computationally expensive operation is finding the projection matrices ${\bf P}_i$ at line 3 of Algorithm \ref{algAPC}. However, when each ${\bf A}_i$ is full rank, the projection matrix ${\bf P}_i={\bf 0}$ and that essentially makes Algorithm \ref{algAPC} a consensus averaging algorithm. Note that we do not need all ${\bf A}_i$s to be full rank, in fact the reason for having the projection matrices ${\bf P}_i$ is to handle the case when ${\bf A}_i$s are not full rank.

In the case where any of ${\bf A}_i$ is not full rank (say when $n_i < M$), we can use the singular value decomposition (SVD) to reduce the computational cost of the steps given at line 2 and line 3 in Algorithm \ref{algAPC}  as listed in Algorithm \ref{AiSVD}.

\begin{algorithm}
  \caption{Using SVD for calculation of ${\bf x}_i^{(0)}$ and ${\bf P}_i$}
\label{AiSVD}
\begin{algorithmic}[1]
  \REQUIRE ${\bf A}_i \in \mathbb{R}^{n_i\times M}$, rank-deficient $n_i <M$, and ${\bf b}_i$.
  \STATE Find SVD ${\bf U}_i {\bf S}_i {\bf V}_i^T={\bf A}_i$ where ${\bf U}_i\in\mathbb{R}^{n_i\times n_i}$, ${\bf V}_i\in\mathbb{R}^{M\times M}$ orthogonal, ${\bf S}_i\in\mathbb{R}^{n_i\times M}$ diagonal.
  \STATE Find ${\bf S}_i^{\dagger}$ by filling its diagonal by the non-zero diagonal terms of ${\bf S}_i$ (at most $n_i$).
  \STATE Find  ${\bf x}_i^{(0)} = {\bf V} {\bf S}_i^{\dagger} {\bf U}^T {\bf b}_i$.
  \STATE Find projection matrix ${\bf P}_i={\bf I}_M - {\bf V}_i \widetilde{\bf I}_M{\bf V}_i^T$ where $\widetilde{\bf I}_M$ is the reduced identity matrix with only the first $n_i$ diagonal elements having $1$.
 \STATE Return ${\bf x}_i^{(0)}$ and ${\bf P}_i$.
\end{algorithmic}
\end{algorithm}

Note that it is possible to make $B$ large such that $n_i=1$ (each sub-matrix is a row) and even in this case, we can get convergence of Algorithm \ref{algAPC}, provided that we select the number of iterations, i.e. $T$, large enough. In section \ref{sec:observations}, we will show an application of Algorithm \ref{algAPC}.

\section{Calibration with diffuse sky models\label{sec:calib}}
In this section, we will expand spectrally and spatially constrained direction dependent calibration to include models for large scale diffuse structure.
Under the assumption that the noise ${\bf N}_{pq}$ in (\ref{V}) has complex, circular Gaussian random variables, the cost function to minimize to get the maximum likelihood estimate of ${\bf J}_{kf_i}$ can be given as
\beqn \label{gcost}
\lefteqn{g_{f_i}(\{{\bf J}_{kf_i},k=0\ldots K\})}\\\nonumber
&&= \sum_{p,q,\Delta_t} \|{\bf V}_{pqf_i}-{\bf J}_{p0f_i} \mathcal{\bf C}_{pqf_i}({\bf S},\widetilde{\bf Z}) {\bf J}_{q0f_i}^H - \sum_{k=1}^K {\bf J}_{pkf_i}{\bf C}_{pqkf_i}{\bf J}_{qkf_i}^H \|^2
\eeqn
where the cost is accumulated over all baselines (that are being part of calibration) and a finite time interval $\Delta_t$. Note that in the case of excluding the short baselines during calibration, the summation is done over a subset of all available baselines.

With the additional constraints (\ref{BZ}) and (\ref{Zbar}), the spectrally and spatially regularized calibration is defined as
\beqn \label{conscalib}
\{{\bf {J}}_{kf_i},\ldots,{\bf {Z}}_k:\ \forall\ k,i\}=\underset{{\bf {J}}_{kf_i},\ldots,{\bf {Z}}_k}{\rm arg\ min} \sum_i g_{f_i}(\{{\bf J}_{kf_i}:\ \forall k\})\\\nonumber
{\rm subject\ to}\ \  {\bf {J}}_{kf_i}={\bf {B}}_{f_i} {\bf {Z}}_k,\ \ i\in[1,P],k\in[0,K]\\\nonumber
{\rm and}\ \ {\bf {Z}}_k=\overline{{\bf Z}}_k, \ \ k\in[0,K].
\eeqn

In conjunction, the update of the global spectral-spatial model can be defined as
\beqn \label{consZ}
\{{\bf Z},\widetilde{\bf Z}\}=\underset{{\bf Z},\widetilde{\bf Z}} {\rm arg\ min} \sum_k \| \overline{{\bf Z}}_k - {\bf Z}{\bmath \Phi}_k\|^2 + \lambda \|{\bf Z}\|^2 + \mu \|{\bf Z}\|_1 \\\nonumber
+ \|\widetilde{\bf Z} - \widetilde{\bf Z}_0\|^2 + \lambda \|\widetilde{\bf Z}\|^2\\\nonumber
{\rm subject\ to}\ \  \widetilde{\bf Z}={\bf Z},
\eeqn
where $\lambda$ ($\in \mathbb{R}^{+}$) is the L-2 regularization factor and $\mu$ ($\in \mathbb{R}^{+}$) is the L-1 regularization factor. Both $\lambda$ and $\mu$ are introduced to keep the global model ${\bf Z}$ low-energy and sparse. Note also that the global model used in $\mathcal{\bf C}_{pqf_i}({\bf S},\widetilde{\bf Z})$ is updated independently of ${\bf Z}$ but under the constraint of them being equal.

In order to solve (\ref{conscalib}), we form the augmented Lagrangian as
\beqn \label{aug}
\lefteqn{
\mathcal{L}(\{{\bf {J}}_{kf_i},{\bf {Z}}_k,{\bf {Y}}_{kf_i},{\bf X}_{k}: \forall\ k,i\}) }\\\nonumber
&& =\sum_i g_{f_i}(\{{\bf J}_{kf_i}:\ \forall k\})\\\nonumber
&& + \sum_{i,k} \left( \| {\bf {Y}}_{kf_i}^H ({\bf {J}}_{kf_i}- {\bf {B}}_{f_i} {\bf {Z}}_k)\| + \frac{\rho}{2} \| {\bf {J}}_{kf_i}- {\bf {B}}_{f_i} {\bf {Z}}_k \|^2 \right) \\\nonumber
&&+\sum_k \left(\|{\bf X}_k^H\left(  {\bf {Z}}_k - \overline{{\bf Z}}_k \right)\|+ \frac{\alpha}{2} \|  {\bf {Z}}_k - \overline{{\bf Z}}_k \|^2 \right)
\eeqn
where the Lagrange multiplies ${\bf Y}_{kf_i}$ ($\in \mathbb{C}^{2N\times 2}$) and ${\bf X}_k$ ($\in \mathbb{C}^{2FN\times 2}$) are introduced for the spectral and spatial constraints. The regularization factors $\rho$ and $\alpha$ ($\in \mathbb{R}^{+}$)  are chosen for each of these constraints. Moreover, it is also possible to select $\rho$ and $\alpha$ distinctly per each direction $k$ (for example, as $\rho_k$ and $\alpha_k$), which is common practice.

Furthermore, in order to solve (\ref{consZ}), we also form the augmented Lagrangian as
\beqn \label{augZ}
\mathcal{L}({\bf Z},\widetilde{\bf Z},{\bmath \Psi}) = \sum_k \| \overline{{\bf Z}}_k - {\bf Z}{\bmath \Phi}_k\|^2 + \lambda \|{\bf Z}\|^2 + \mu \|{\bf Z}\|_1 \\\nonumber 
+ \|\widetilde{\bf Z} - \widetilde{\bf Z}_0\|^2 + \lambda \|\widetilde{\bf Z}\|^2\\\nonumber
+ \|{\bmath \Psi}^H({\bf Z}-\widetilde{\bf Z})\| +\frac{\gamma}{2}\|{\bf Z}-\widetilde{\bf Z}\|^2
\eeqn
where the Lagrange multiplier ${\bmath \Psi}$ ($\in \mathbb{C}^{2FN \times 2G}$) and the corresponding regularization factor $\gamma$ ($\in \mathbb{R}^{+}$) are used for the constraint (\ref{Ztildecons}).

Minimizing both (\ref{aug}) and (\ref{augZ}) can be achieved as two interleaved iterations of the alternating direction method of multipliers (ADMM) algorithm \citep{boyd2011}. The essence of the minimization can be summarized as
\begin{enumerate}
  \item \label{step1} Solve ${\bf {J}}_{kf_i}$ using generic direction dependent calibration \citep{Kaz2,Kaz3} to minimize (\ref{aug}).
  \item \label{step2} Solve ${\bf {Z}}_k$ by setting gradient of $\mathcal{L}(\{{\bf {J}}_{kf_i},{\bf {Z}}_k,{\bf {Y}}_{kf_i},{\bf X}_{k}: \forall\ k,i\})$ to zero as in (\ref{zsol}).
  \item \label{step3} Solve for ${\bf Z}$ and $\widetilde{\bf Z}$ by minimizing (\ref{augZ}).
  \item \label{step4} Update prediction of diffuse sky model using $\widetilde{\bf Z}$.
\end{enumerate}
By repeating the steps \ref{step1}-\ref{step4} above, we are able to reach a stable calibration end result as we show using simulations and real data in sections \ref{sec:results} and \ref{sec:observations}.

We elaborate some of the above steps \ref{step1}-\ref{step4} in the following text. In step \ref{step2}, the solution for ${\bf Z}_k$ is obtained in closed form as
\beqn
\label{zsol}
\lefteqn{{\bf {Z}}_k= }\\\nonumber
&& \left( \sum_i \rho {\bf {B}}_{f_i}^H {\bf {B}}_{f_i} + \alpha {\bf I}_{2FN} \right)^{\dagger}
\left(\sum_i {\bf {B}}_{f_i}^H \left({\bf {Y}}_{kf_i} + \rho {\bf {J}}_{kf_i}\right) + \alpha  \overline{{\bf Z}}_k -{\bf X}_{k} \right).
\eeqn

In step \ref{step3}, solving for ${\bf Z}$ involves a sparsity constrained minimization, hence we use the fast iterative shrinkage and thresholding algorithm (FISTA \cite{FISTA}) to find ${\bf Z}$. In order to use FISTA, we need to provide the differentiable cost function $h({\bf Z})$,
\beq \label{hh}
h({\bf Z})=\sum_k \|\overline{{\bf Z}}_k - {\bf Z} {\bmath \Phi}_k \|^2 + \lambda \|{\bf Z}\|^2 +\|{\bmath \Psi}^H({\bf Z}-\widetilde{\bf Z})\| +\frac{\gamma}{2}\|{\bf Z}-\widetilde{\bf Z}\|^2
\eeq
and the gradient
\beq \label{gradh}
\nabla h = {\bf Z}\left( \sum_k {\bmath \Phi}_k {{\bmath \Phi}_k}^H + \lambda {\bf I}_{2G} \right) - \sum_k \overline{{\bf Z}}_k {{\bmath \Phi}_k}^H + \frac{1}{2} {\bmath \Psi} + \frac{\gamma}{2}({\bf Z}-\widetilde{\bf Z}).
\eeq

Once we have found ${\bf Z}$ in (\ref{augZ}), we find $\widetilde{\bf Z}$ in closed form by finding the gradient of (\ref{augZ}),
\beq \label{gradZ}
{\rm grad}_{\widetilde{\bf Z}}\left(\mathcal{L}({\bf Z},\widetilde{\bf Z},{\bmath \Psi}) \right) = \widetilde{\bf Z} - \widetilde{\bf Z}_0 - \frac{1}{2} {\bmath \Psi} - \frac{\gamma}{2}({\bf Z}-\widetilde{\bf Z}) + \lambda \widetilde{\bf Z}
\eeq
and equating it to zero as,
\beq \label{Ztilde}
\widetilde{\bf Z} =\frac{1}{1+\frac{\gamma}{2}+\lambda} \left(\widetilde{\bf Z}_0 + \frac{1}{2} {\bmath \Psi} + \frac{\gamma}{2} {\bf Z} \right).
\eeq

The complete calibration algorithm with the inclusion of the diffuse sky model is summarized in Algorithm \ref{algSADMM}. This algorithm is designed to run on a distributed computer with one fusion center and several compute agents. In this manner we maximize the parallelism by exploiting the inherent data parallelism of radio interferometers (that collect and store data at multiple frequencies).

\begin{algorithm}
  \caption{Spectrally and spatially constrained distributed calibration with diffuse sky models}
\label{algSADMM}
\begin{algorithmic}[1]
\REQUIRE $A$: number of ADMM iterations, $C$: cadence of spatial model update, $\rho,\alpha$,$\gamma$: regularization factors.
  \STATE Initialize ${\bf {Y}}_{kf_i}$,$\overline{{\bf Z}}_k$,${\bf X}_k$,${\bmath \Psi}$ to zero $\forall k,i$.
  \STATE Initialize $\widetilde{\bf Z} \leftarrow \widetilde{\bf Z}_0$ using (\ref{Z0}).
\FOR{$a=1,\ldots,A$}
\STATE \COMMENT{Do in parallel $\forall k,i$ at all compute agents and the fusion center}
 \STATE Compute agents minimize (\ref{aug}) for ${\bf {J}}_{kf_i}$.
  \STATE Fusion center solves (\ref{zsol}) for ${\bf Z}_k$.
 \STATE ${\bf {Y}}_{kf_i} \leftarrow {\bf {Y}}_{kf_i} + \rho \left({\bf {J}}_{kf_i}-{\bf {B}}_{f_i} {\bf {Z}}_k\right)$.
  \IF{$a$ is a multiple of $C$}
  \STATE Fusion center updates ${\bf Z}$ using (\ref{hh}) and (\ref{gradh}).
  \STATE Fusion center updates $\overline{{\bf Z}}_k$ using (\ref{Zbar}).
\STATE ${\bf X}_k \leftarrow {\bf X}_k + \alpha \left(  {\bf {Z}}_k - \overline{{\bf Z}}_k\right)$.
  \STATE Fusion center updates $\widetilde{\bf Z}$ using (\ref{Ztilde}).
  \STATE ${\bmath \Psi} \leftarrow {\bmath \Psi} + \gamma({\bf Z}-\widetilde{\bf Z})$.
  \STATE Compute agents update diffuse sky model prediction $\mathcal{\bf C}_{pqf_i}({\bf S},\widetilde{\bf Z})$ $\forall i$.
 \ENDIF
\ENDFOR
\end{algorithmic}
\end{algorithm}

\section{Simulation results\label{sec:results}}
In this section, we present results to illustrate the efficacy of the proposed Algorithm \ref{algSADMM} based on simulated observations. The simulations are done in a Monte Carlo manner, where we randomize the sky model (including the large scale diffuse structure) as well as the systematic errors in addition to the noise. We simulate the LOFAR highband antenna (HBA) array \citep{LOFAR}, with $N=62$ stations observing in the frequency range $[115,185]$ MHz. The pointing center of the telescope is randomly selected to be either the north celestial pole (NCP) or 3C196. We select $P=8$ channels uniformly covering the frequency range and the duration of each observation is kept at $10$ min, with $10$ s integration time.

The simulated sky model consists of $K=20$ compact sources, randomly spread across a $30\times 30$ square degrees field of view. Their intrinsic flux densities are randomly chosen in the range $[1,10]$ Jy and their spectral indices are randomly sampled from a standard normal distribution. An additional $400$ weak sources (point sources and Gaussian sources), with flux densities in the range $[0.01,0.5]$ Jy and a power law spectral index of $-2$ are randomly placed in the field of view and are not included in the calibration sky model.

The diffuse sky model is randomly generated as follows. In each simulation, we generate three shapelet models for the Stokes I, Q and U diffuse emission. In each shapelet model, the model order $\sqrt{M}$ is uniformly sampled from $[10,20]$, therefore each shapelet decomposition will have $M$ basis functions. The scale of the shapelet basis $\beta$ is uniformly sampled from the range $[0.01,1]$ but if this is too high for the field of view, i.e., $\sqrt{M}\times \beta > 2$, we sample $\beta$ from the range $[2,2.001]/\sqrt{M}$. Finally, the $M$ shapelet coefficients are randomly sampled from the standard normal distribution. We only use these shapelet models for simulation of data, and when we calibrate, we perturb the shapelet models to make it more error prone and realistic (we do not use Algorithm \ref{algAPC} in this section). We perturb the scale $\beta$ and the coefficients by 10\% of their original value by adding appropriately scaled noise drawn from the standard normal distribution.

The systematic errors ${\bf J}_{kf_i}$ in (\ref{Jf}) are generated to have specific properties, i.e., randomness, smooth variation over time and frequency, and smoothness over space (over $k$). We first generate $8N$ random planes in space having their domain being equal to the image coordinates. The $8N$ planes are for each real value of ${\bf J}_{kf_i}$. For a selected $f_i$, we fill ${\bf J}_{kf_i}$  with values drawn from the complex standard normal distribution and add a fraction of the value of the random planes evaluated at the coordinates for direction $k=0,\ldots,K$. Next, we generate $8N$ third order random polynomials in frequency, and multiply ${\bf J}_{kf_i}$ to evaluate the systematic errors for all frequencies. Simultaneously, we generate $8N$ random sinusoidal polynomials in time and multiply ${\bf J}_{kf_i}$ with these polynomials to evolve the systematic errors over time. During simulation, the systematic errors are used to corrupt the data. In addition, the LOFAR HBA beam model (station beam and dipole beam) are also applied to the data. Note that when simulating the diffuse shapelet models, we only apply the systematic errors evaluated at the center of the source.

We add noise to the simulated data with a signal to noise ratio of $1/0.05$, the elements of ${\bf N}_{pq}$ in (\ref{V}) are drawn from a complex standard normal distribution. 

During spectrally and spatially regularized calibration, we use a Bernstein polynomial basis with $F=3$ terms for ${\bf b}_{f_i}$ and a shapelet basis with $G=4$ terms for ${\bmath \phi}_k$ in (\ref{bases}). The scale of the shapelet basis is automatically determined depending on the spread of the $K$ directions given in the sky model. We evaluate three calibration settings:
\begin{enumerate}
\item Calibration with only spectral regularization.
\item Calibration with both spectral and spatial regularization.
\item Calibration with both spectral and spatial regularization with a diffuse sky model.
\end{enumerate}

The regularization factors $\rho_k$ and $\alpha_k$ in (\ref{aug}) are determined as follows. We set $\rho_k$ proportional to the flux density of the $k$-th source, with the maximum value of $\rho_k$ being limited to $200$. We set $\alpha_k$ inversely proportional to the distance of the $k$-th source from the pointing center. We limit the maximum value of $\alpha_k$ to $1$. When the diffuse foreground model is also included in calibration, we set $\rho=1$, $\alpha=0.1$ and $\gamma=0.1$ for this direction (in addition, $\lambda=\mu=1e-3$). The number of ADMM iterations is set to $A=50$ and the cadence of spatial model update is set to $C=3$.

The calibration is repeated for every $10$ time samples, or in other words, for every $10\times 10=100$ s. Note that we do not exclude the short baselines during calibration. Therefore, all baselines are used to evaluate (\ref{gcost}). Once the solutions $\widehat{\bf J}_{kf_i}$ are found, the residual is calculated as
\beq \label{residual}
 {\bf R}_{pqf_i} = {\bf V}_{pqf_i} - \sum_{k=1}^K \widehat{\bf J}_{pkf_i}{\bf C}_{pqkf_i}\widehat{\bf J}_{qkf_i}^H.
\eeq
Note that we do not subtract the diffuse sky model in (\ref{residual}) and it should remain in the residual.

\begin{figure*}
\begin{minipage}{0.98\linewidth}
\begin{center}
\begin{minipage}{0.98\linewidth}
\centering
  \centerline{\includegraphics[width=0.5\textwidth]{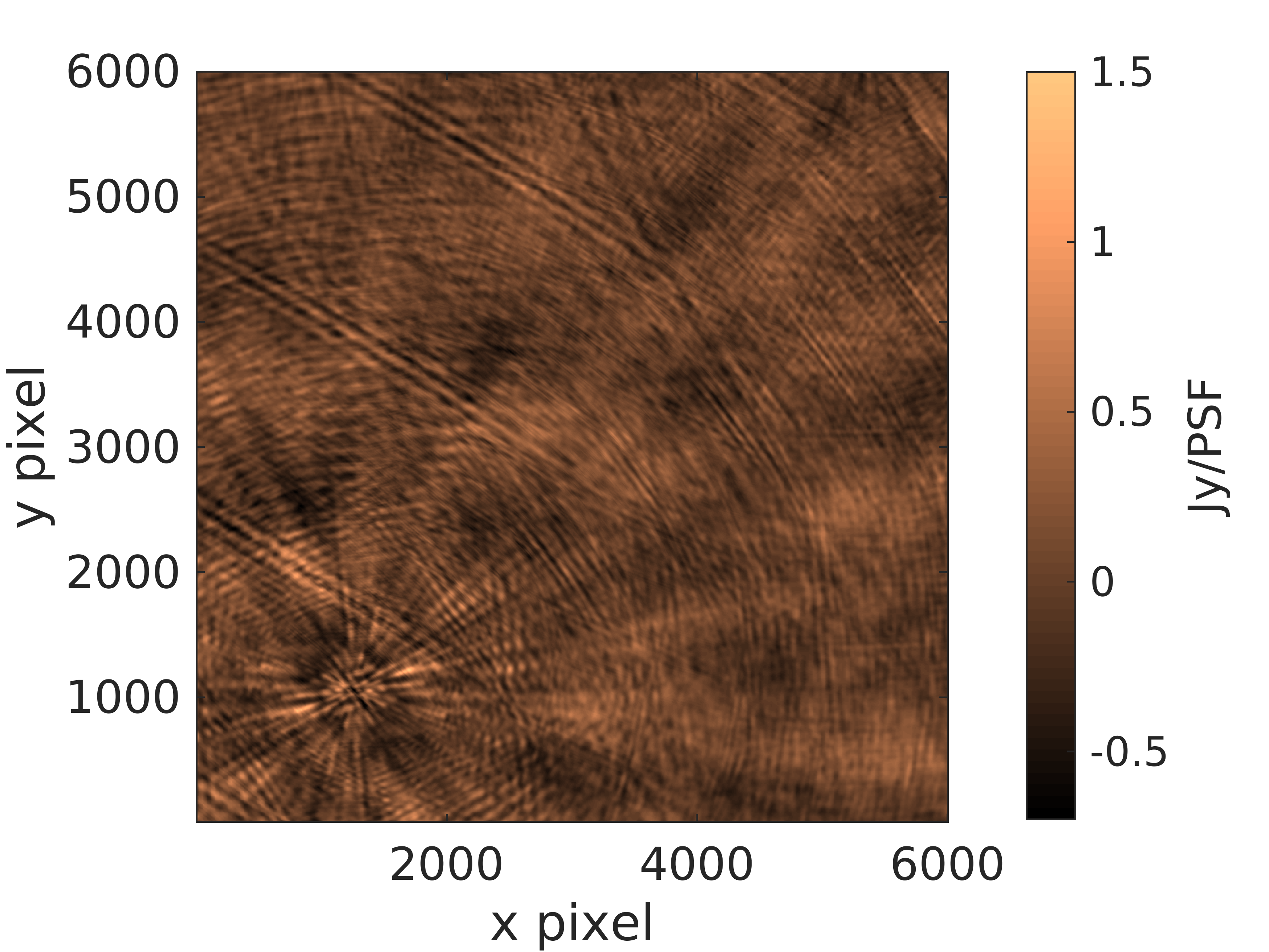}}
\vspace{0.5cm} \centerline{(a)}\smallskip
\end{minipage}\\
\begin{minipage}{0.48\linewidth}
\centering
 \centerline{\includegraphics[width=1.0\textwidth]{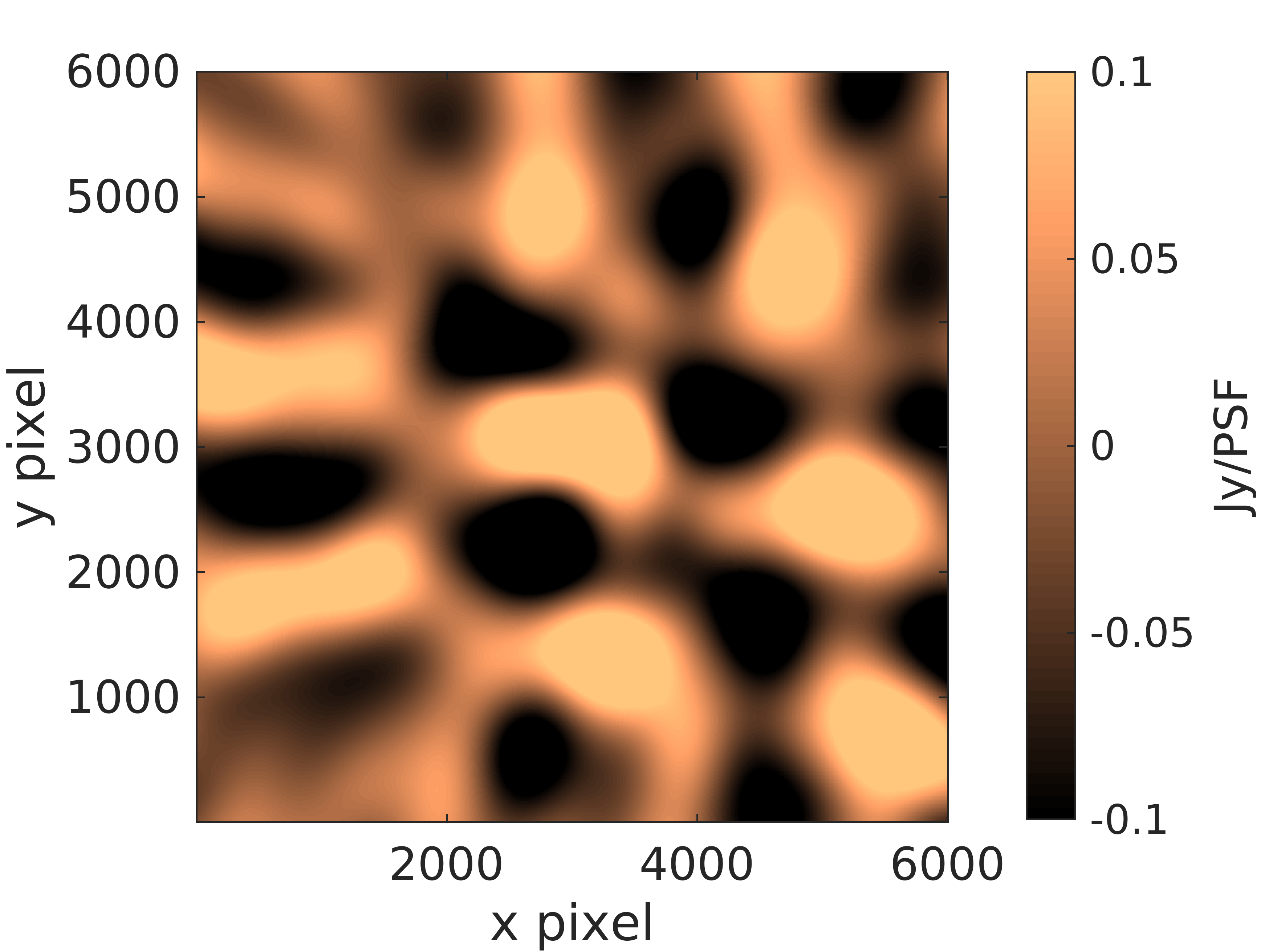}}
\vspace{0.5cm} \centerline{(b)}\smallskip
\end{minipage}
\begin{minipage}{0.48\linewidth}
\centering
 \centerline{\includegraphics[width=1.0\textwidth]{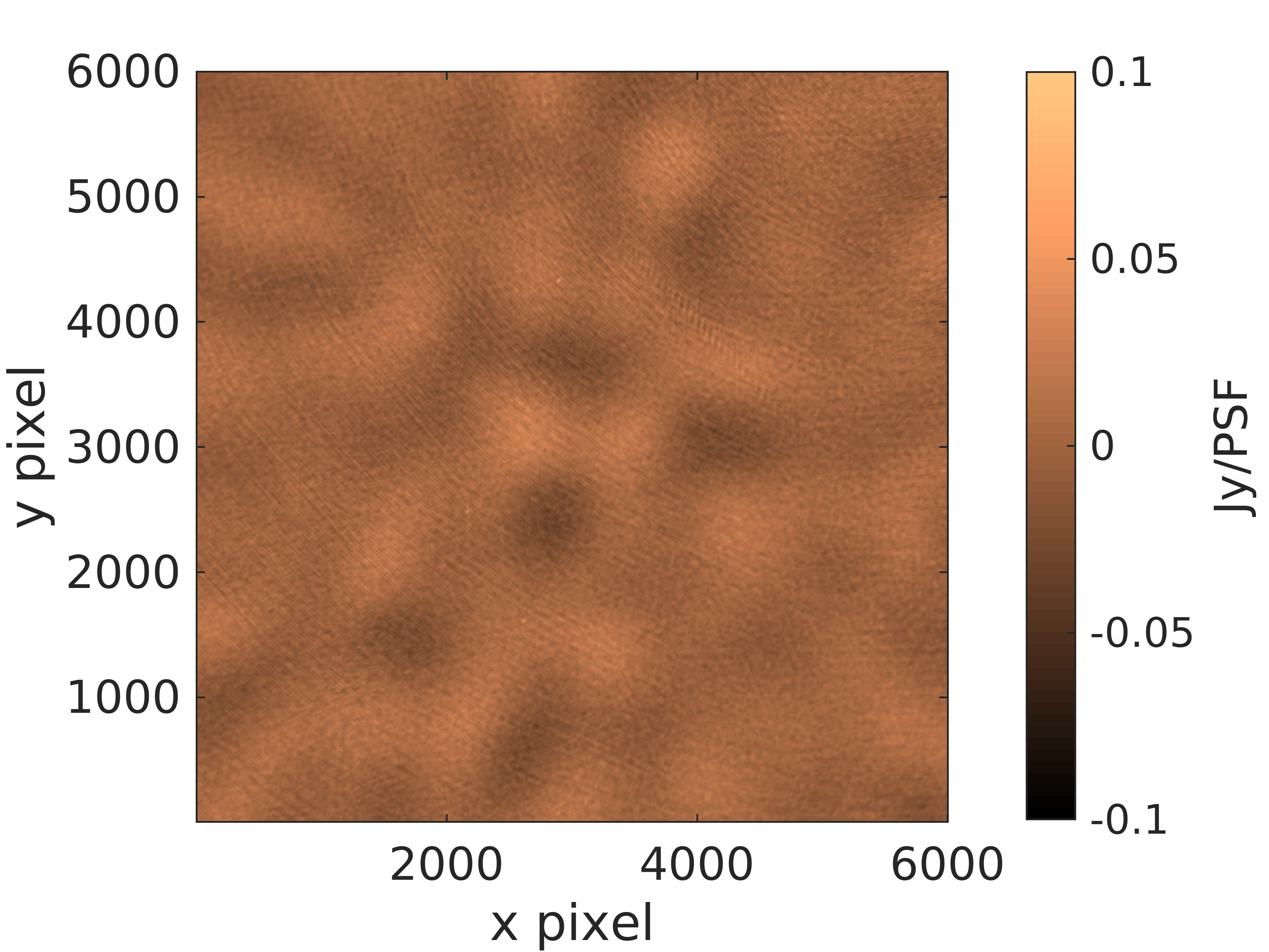}}
\vspace{0.5cm} \centerline{(c)}\smallskip
\end{minipage}\\
\begin{minipage}{0.48\linewidth}
\centering
 \centerline{\includegraphics[width=1.0\textwidth]{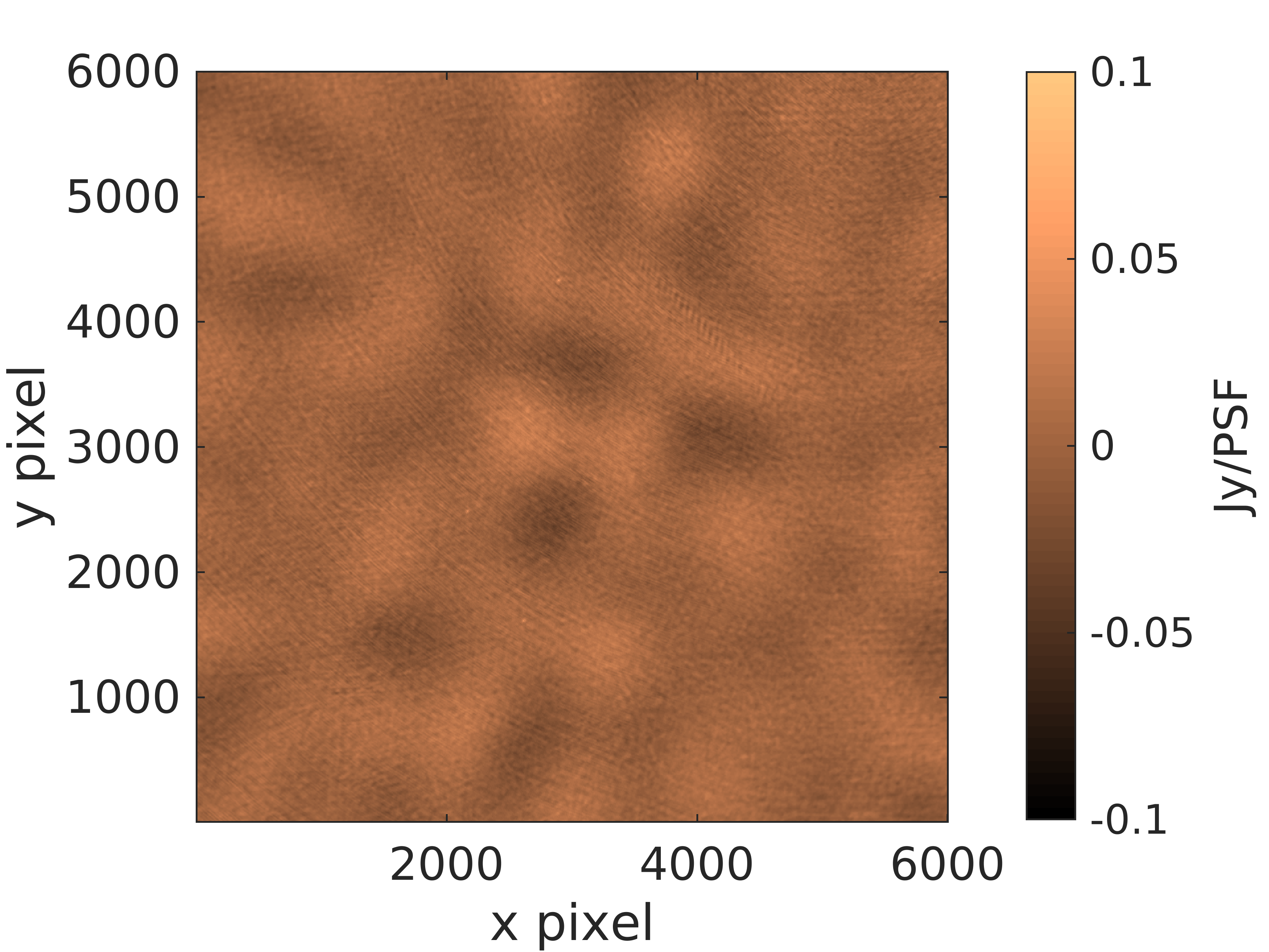}}
\vspace{0.5cm} \centerline{(d)}\smallskip
\end{minipage}
\begin{minipage}{0.48\linewidth}
\centering
 \centerline{\includegraphics[width=1.0\textwidth]{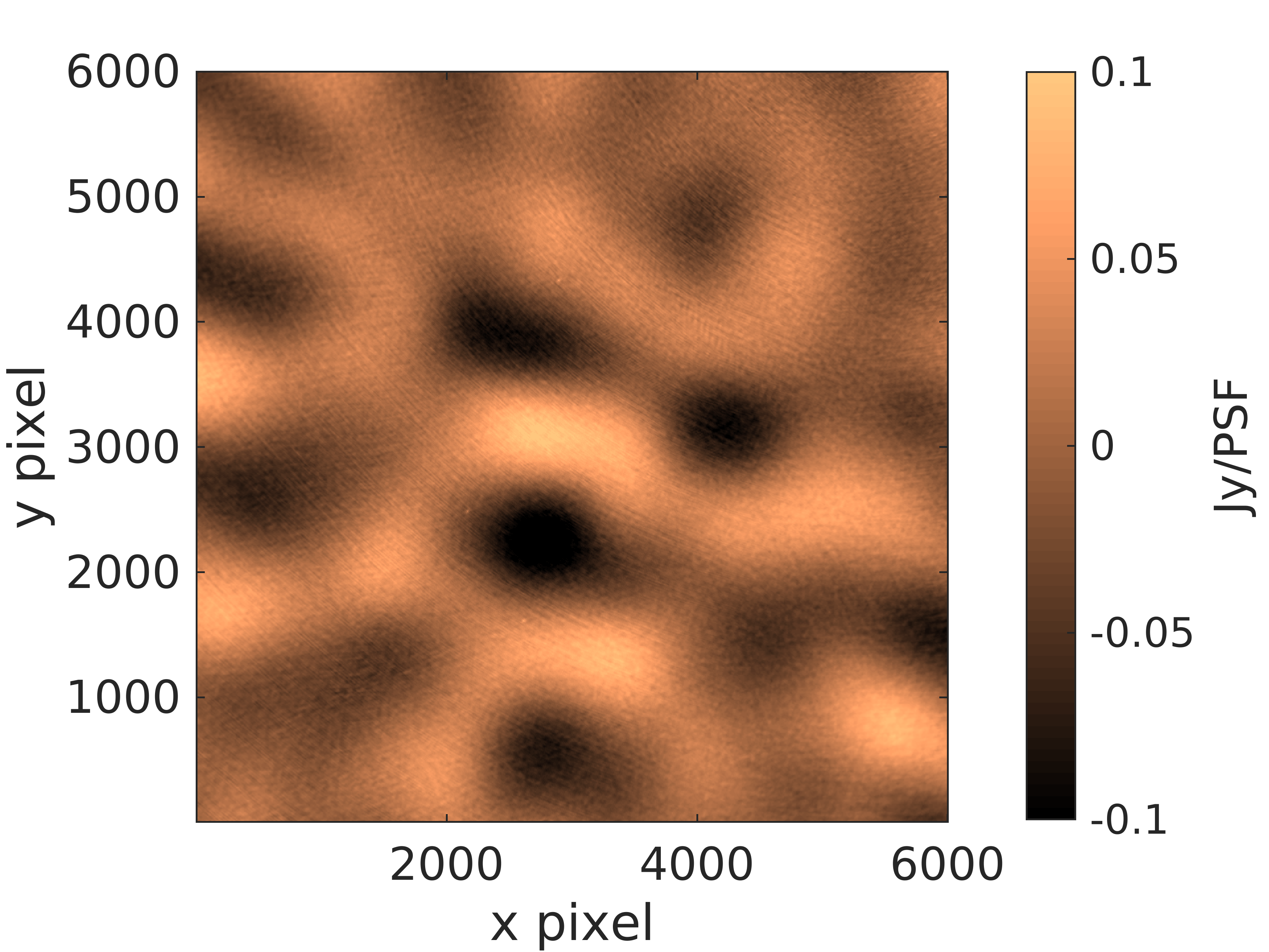}}
\vspace{0.5cm} \centerline{(e)}\smallskip
\end{minipage}
\end{center}
\caption{
  Sample images (Stokes I, not deconvolved, natural weights) made using simulated data, covering about $10 \times 10$ square degrees in the sky. (a) The image before calibration. (b) The diffuse sky and the weak sources that are hidden in the simulated data. (c) The residual image after calibration only using spectral regularization. (d) The residual image after calibration with spectral and spatial regularization. (e)  The residual image after calibration with spectral and spatial regularization and using the diffuse sky model. Images (b)-(e) have the same intensity scale.
\label{simulated_maps}}
\end{minipage}
\end{figure*}

We make images of the residual data using natural weights to enhance the large scale structure (if any). One such example is shown in Fig. \ref{simulated_maps}. In addition to showing the data and residual images, we also show the image where we only simulate the diffuse model in Fig. \ref{simulated_maps} (b). It is obvious that the residual image obtained after calibration with a diffuse sky model in Fig. \ref{simulated_maps} (e) agrees well with the diffuse model. In contrast, Figs. \ref{simulated_maps} (c) and (d) shows much more suppression of the diffuse model, which is the main reason to remedy this issue, for example by excluding short baselines in calibration. 

\begin{figure}
\begin{minipage}{0.98\linewidth}
\begin{center}
\begin{minipage}{0.98\linewidth}
\centering
  \centerline{\includegraphics[width=1.0\textwidth]{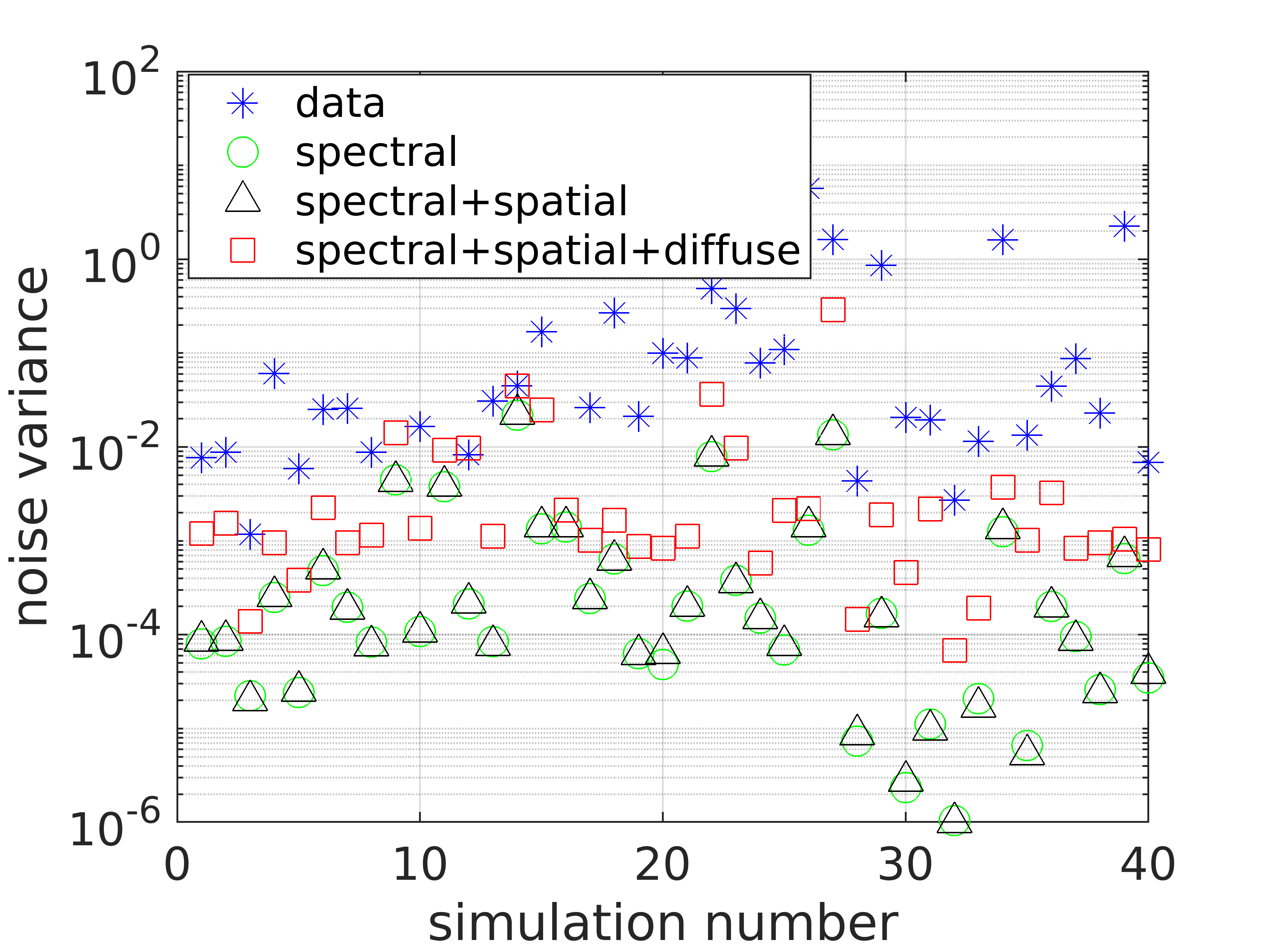}}
\vspace{0.5cm} \centerline{(a)}\smallskip
\end{minipage}\\
\begin{minipage}{0.98\linewidth}
\centering
 \centerline{\includegraphics[width=1.0\textwidth]{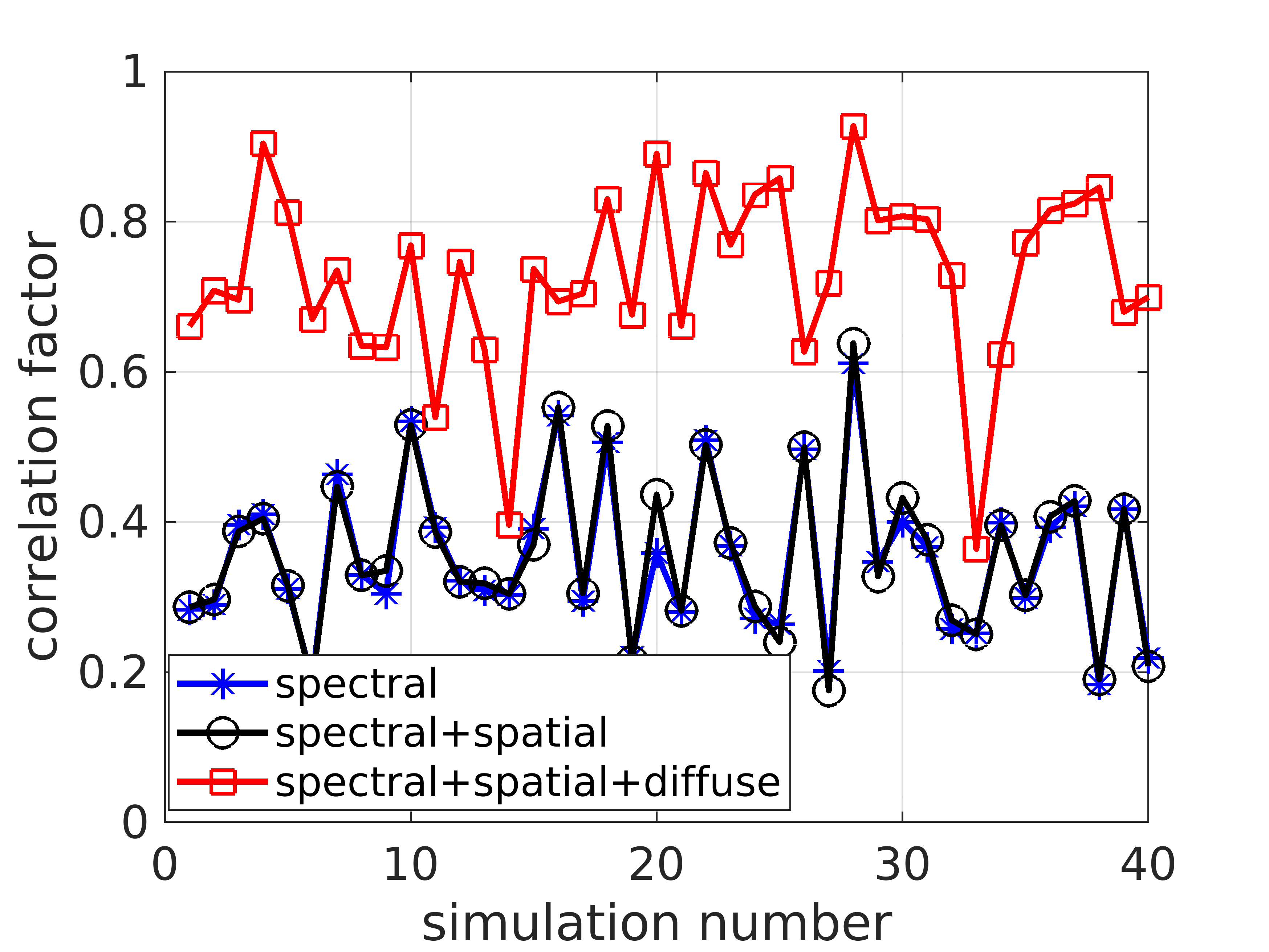}}
\vspace{0.5cm} \centerline{(b)}\smallskip
\end{minipage}
\end{center}
\caption{
  Results of $40$ simulations. (a) The variance of images made before and after calibration. Three calibration scenarios are shown where calibration without using a diffuse sky model gives lower noise, probably due to overfitting. (b) The correlation of the diffuse foreground with the residual. Including the diffuse sky model in calibration preserves much of the diffuse sky model. However, in other calibration scenarios, we see significant suppression of the diffuse sky (hence the reason for excluding short baselines in such calibration scenarios).
\label{noise_corr}}
\end{minipage}
\end{figure}

We perform $40$ Monte Carlo simulations and we provide some results in Fig. \ref{noise_corr}. To evaluate the simulations quantitatively, we provide two metrics. First, we measure the variance (calculated using the whole image) of images similar to the ones shown in Fig. \ref{simulated_maps}. We expect the variance in the residual to decrease as we subtract most of the signal to calculate the residual (\ref{residual}). However, we should not expect the noise variance to go too-low, indicating the presence of some overfitting. The number of constraints compared to the degrees of freedom used is more or less on the border of becoming ill-constrained for our calibration setup and this will be reflected in noise reduction below a value that is realistic. As shown in Fig. \ref{noise_corr} (a), we see that the noise variance of the residual images for calibration without using the diffuse sky model is generally lower than the variance of the residual images that have been calibrated using a diffuse sky model.

A second metric is a measure of the diffuse foreground power remaining in the residual. In order to measure this, we find the correlation factor between the image of the diffuse foreground (such as Fig. \ref{simulated_maps} (b)) and the residual image. A higher correlation  will indicate lower suppression of the diffuse foreground in the residual image. Since we work with data in full polarization, we find the correlation factor of Stokes I images, Stokes Q images and Stokes U images separately and find the mean absolute value of the correlation. In Fig. \ref{noise_corr} (b), we compare the correlation for each of the calibration settings. Once again, we see that calibration without using a diffuse sky model leads to significant suppression of the diffuse foreground while using a diffuse sky model (even an approximate one) will remedy this issue.

\section{Real observations\label{sec:observations}}
In this section, we provide an example based on real data that includes substantial diffuse emission. We consider an observation of the galactic plane (centered at the supernova remnant G46.08-0.3) using the LOFAR HBA array. We use 100 subbands in the frequency range $[124,148]$ MHz with each subband having a bandwidth of $0.192$ MHz. We use a 4 hour duration of data that were acquired by the project LC4\_010 in July 2015 \citep{polderman2021depth} and are now public at the LOFAR long term archive.

\begin{figure*}
\begin{minipage}{0.98\linewidth}
\begin{center}
\begin{minipage}{0.98\linewidth}
\centering
  \centerline{\includegraphics[width=0.7\textwidth]{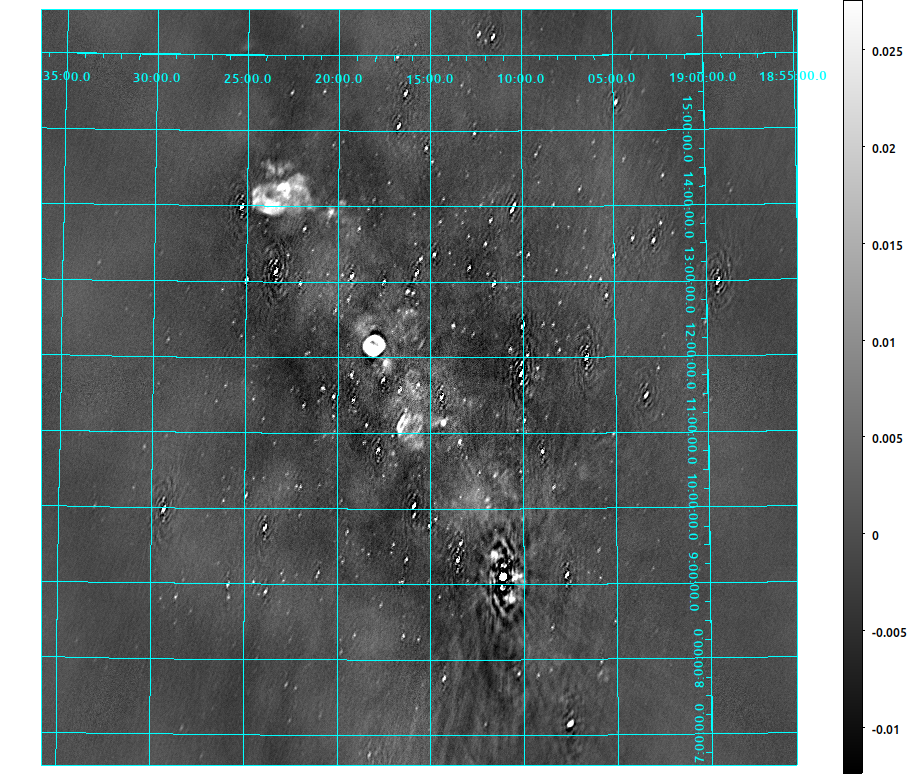}}
\vspace{0.5cm} \centerline{(a)}\smallskip
\end{minipage}\\
\begin{minipage}{0.98\linewidth}
\centering
 \centerline{\includegraphics[width=0.7\textwidth]{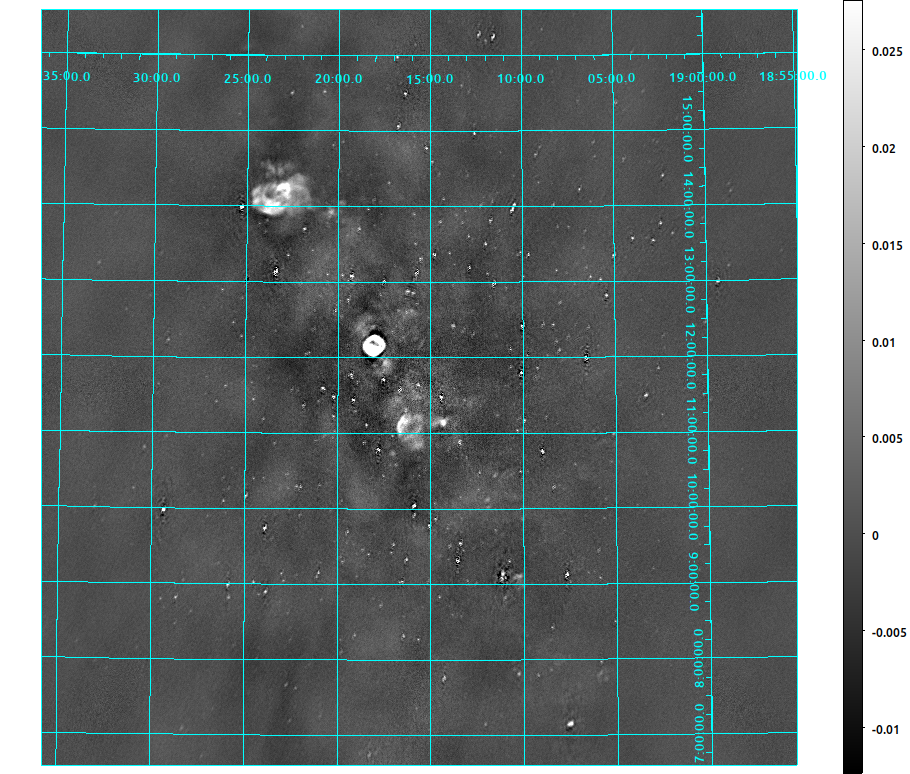}}
\vspace{0.5cm} \centerline{(b)}\smallskip
\end{minipage}
\end{center}
\caption{
  Images (a) before, and (b) after direction dependent calibration, at 137 MHz after averaging all images made by 100 subbands. The images are not deconvolved and are of size  $18000 \times 18000$ pixels with $2^{\prime\prime}$ pixel size. The units of the colourbar are in Jy/PSF. The grid shows right ascension and declination.
\label{full_g46}}
\end{minipage}
\end{figure*}

The image after direction independent calibration and averaging all 100 subbands is shown in Fig. \ref{full_g46} (a). All the images shown are not deconvolved (i.e., dirty images). This image shows a field of view of about $10\times 10$ square degrees using $18 000 \times 18000$ pixels of size $2^{\prime\prime}$.

\begin{figure*}
\begin{minipage}{0.98\linewidth}
\begin{center}
\begin{minipage}{0.98\linewidth}
\centering
  \centerline{\includegraphics[width=0.7\textwidth]{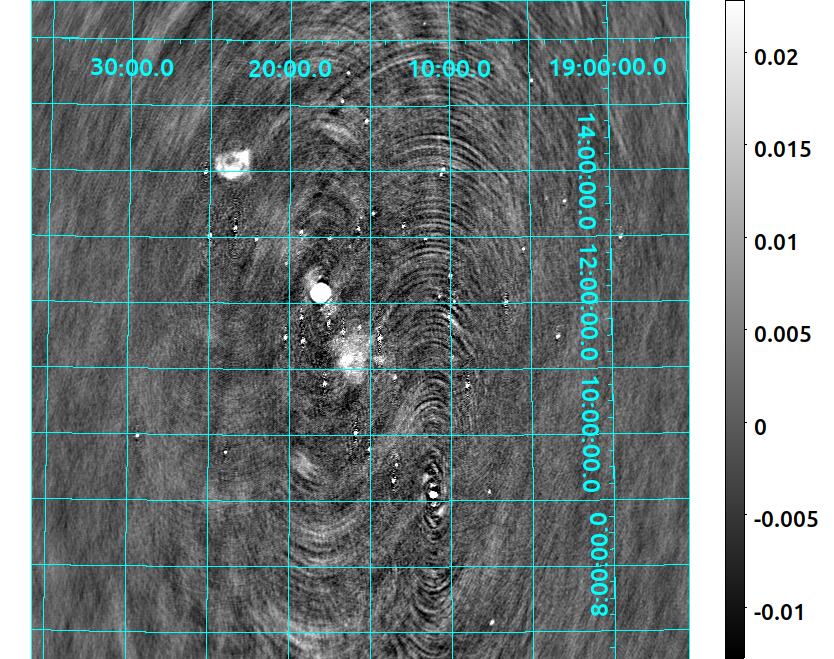}}
\vspace{0.5cm} \centerline{(a)}\smallskip
\end{minipage}\\
\begin{minipage}{0.98\linewidth}
\centering
 \centerline{\includegraphics[width=0.7\textwidth]{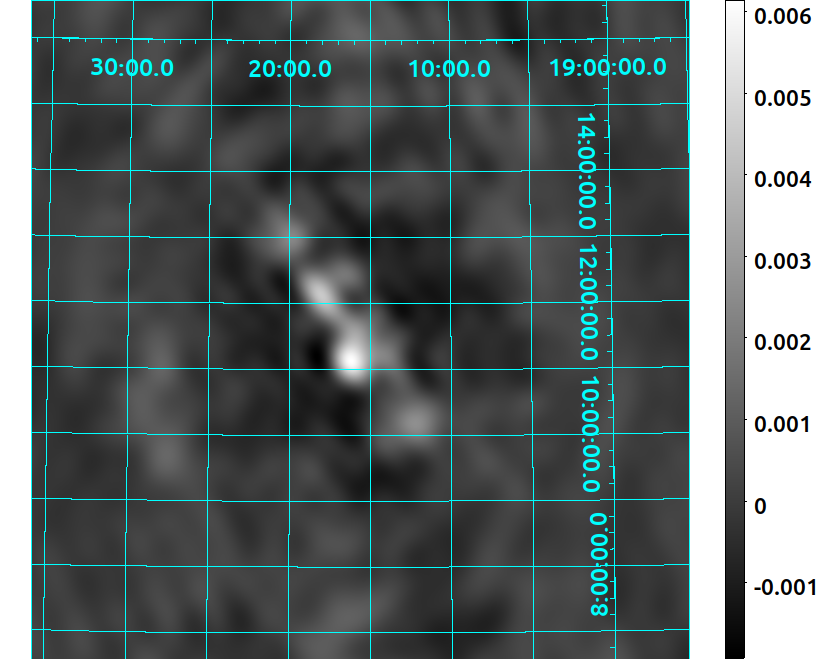}}
\vspace{0.5cm} \centerline{(b)}\smallskip
\end{minipage}
\end{center}
\caption{
  The sky model simulated for a single subband at 137 MHz. The image size is $18000 \times 18000$ pixels with $2^{\prime\prime}$ pixel size. The full sky model is shown in (a) and the large scale diffuse sky model is shown in (b). The shapelet model used in (b) has $M=400$ basis functions. The colourbar units are in Jy/PSF but note the difference in the colourbar limits in (a) and (b). 
\label{model_g46}}
\end{minipage}
\end{figure*}

For direction dependent calibration, we build a sky model of about 100 compact point sources and four shapelet models to represent the compact diffuse structure using the data that made the image in Fig. \ref{full_g46} (a). In addition, we create one shapelet model for the large scale diffuse background using Algorithm \ref{algAPC} as described in section \ref{sec:apc} (the total number of pixels $N_I$ is over 300 million requiring its use here). In Fig. \ref{model_g46}, we show the full sky model (simulated at a single frequency at 137 MHz) as well as the model for the large scale diffuse structure. Note that the shapelet models are not 100\% accurate as seen by comparing Figs. \ref{full_g46} and \ref{model_g46}. We intentionally kept this discrepancy to enhance the effect due to model error. Note also that the image in Fig. \ref{full_g46} (a) shows more compact sources than what is included in our model shown in Fig. \ref{model_g46} (a). Indeed, we have only included about $1/3$ of all the visible point sources in the sky model and this also is to test our algorithm under less favorable circumstances.  

We group the sky model into 30 source clusters ($K=30$) for direction dependent calibration with a solution time interval of 5 min. We consider three calibration scenarios:
\begin{enumerate}
  \item Calibration without including the large scale diffuse sky model and using all baselines (excluding autocorrelations). Spectral regularization with $F=3$.
  \item Calibration without including the large scale diffuse sky model, but excluding baselines shorter than 300 wavelengths from calibration. Spectral regularization with $F=3$. The residual is calculated for all baselines.
  \item Calibration with the inclusion of a large scale diffuse sky model (Algorithm \ref{algSADMM}) and using all baselines excluding autocorrelations. Spectral and spatial regularization with $F=3$, $G=4$, $\gamma=0.1$.
\end{enumerate}

\begin{figure*}
\begin{minipage}{0.98\linewidth}
\begin{center}
\begin{minipage}{0.98\linewidth}
\centering
  \centerline{\includegraphics[width=0.4\textwidth]{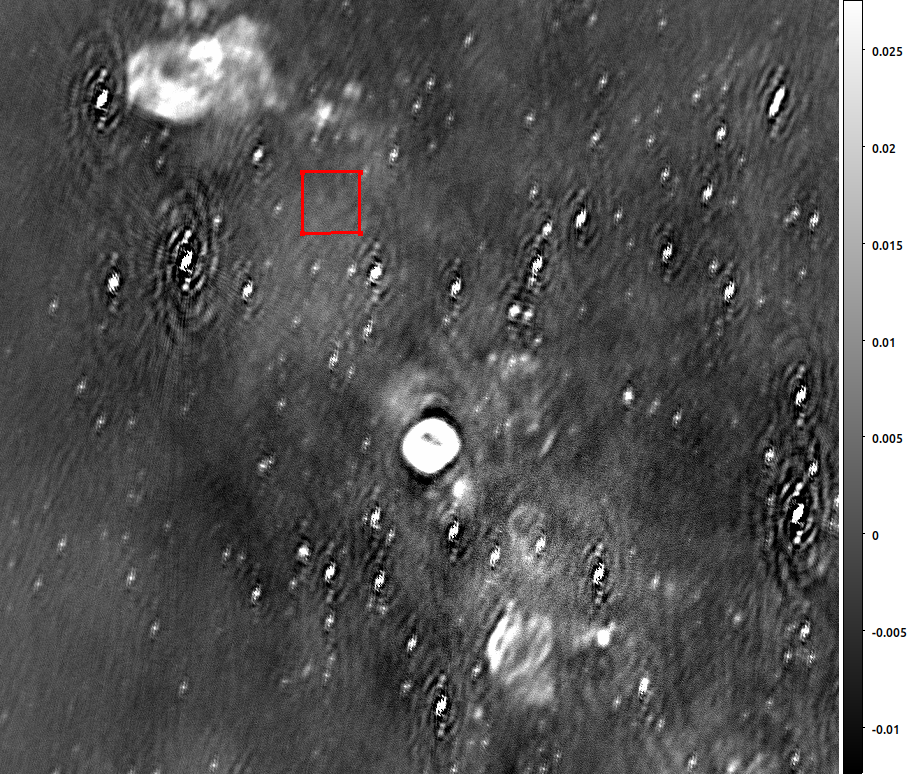}}
\vspace{0.5cm} \centerline{(a)}\smallskip
\end{minipage}\\
\begin{minipage}{0.48\linewidth}
\centering
 \centerline{\includegraphics[width=0.8\textwidth]{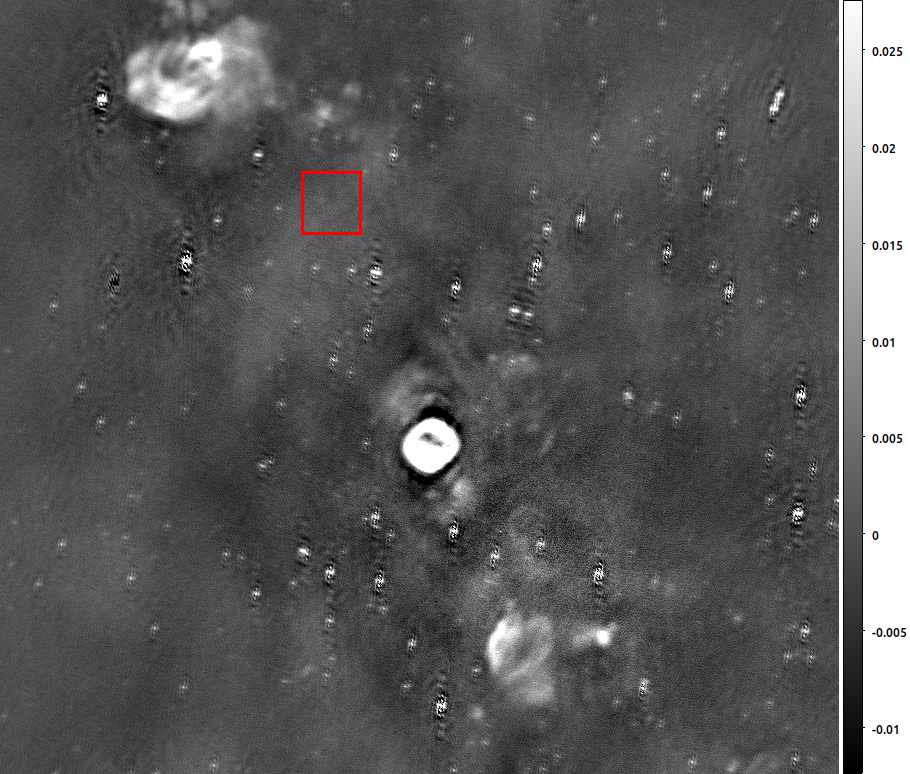}}
\vspace{0.5cm} \centerline{(b)}\smallskip
\end{minipage}
\begin{minipage}{0.48\linewidth}
\centering
 \centerline{\includegraphics[width=0.8\textwidth]{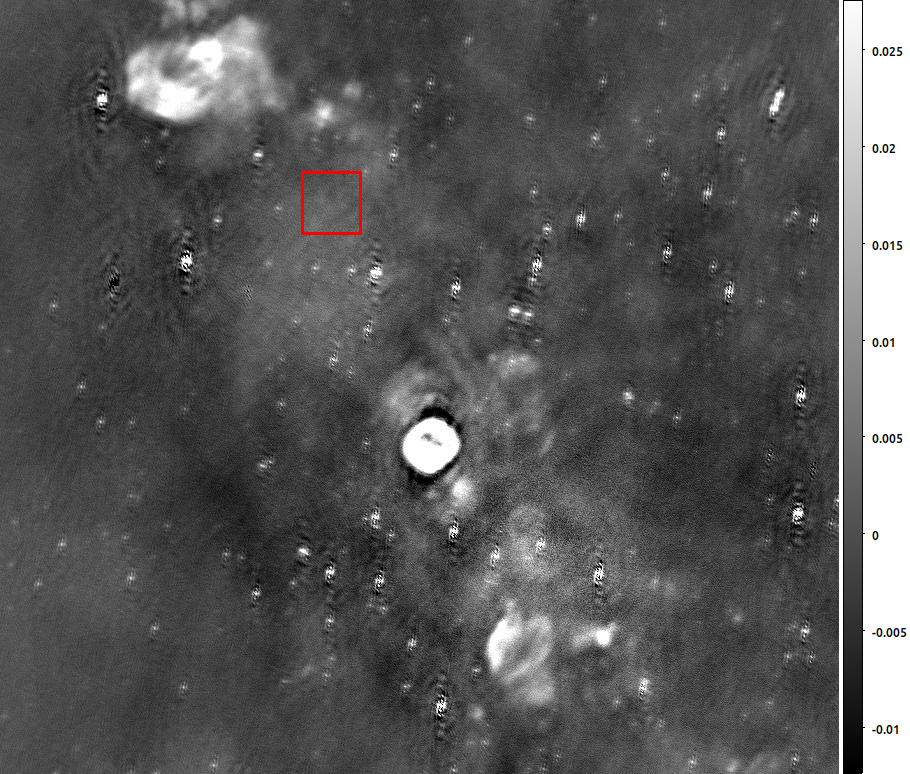}}
\vspace{0.5cm} \centerline{(c)}\smallskip
\end{minipage}\\
\begin{minipage}{0.98\linewidth}
\centering
 \centerline{\includegraphics[width=0.4\textwidth]{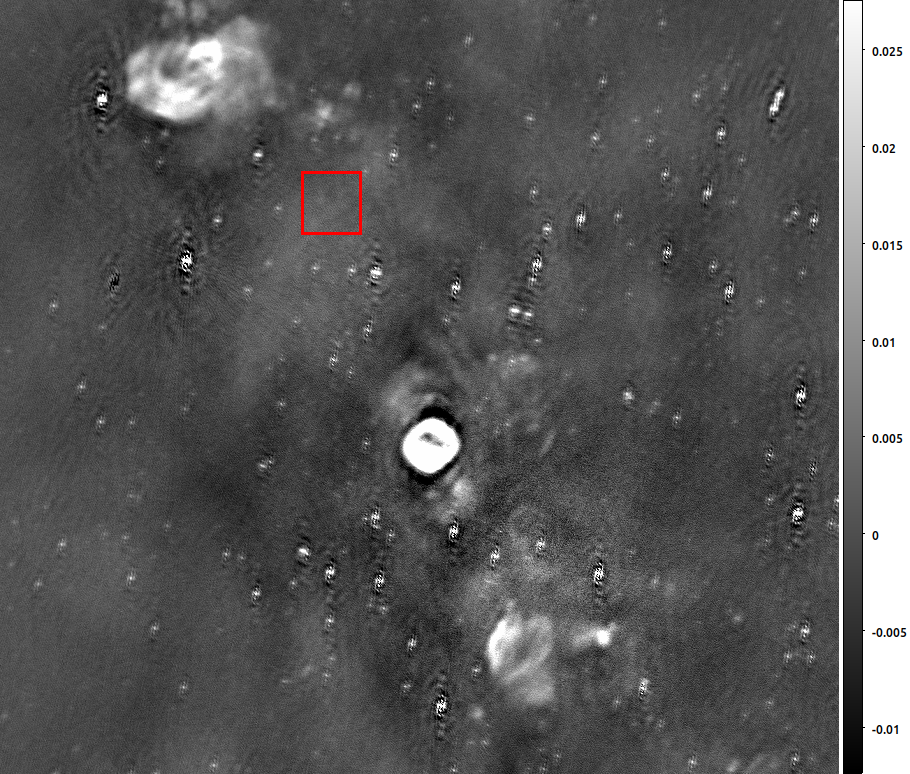}}
\vspace{0.5cm} \centerline{(d)}\smallskip
\end{minipage}
\end{center}
\caption{
  Zoomed in image covering about $4\times 4$ square degrees, (a) before calibration (b) calibration with all baselines (c) calibration excluding baselines shorter than 300 wavelengths, and (d) calibration without excluding short baselines and with a model for the large scale diffuse structure. The rectangle in red is about  $500 \times 500$ pixels and within this rectangle, the sum of all pixels are (a) 1105 (b) 597 (c) 1154 and (d) 977 Jy/PSF, respectively.
\label{zoom_g46}}
\end{minipage}
\end{figure*}

\begin{figure*}
\begin{minipage}{0.98\linewidth}
\begin{center}
\begin{minipage}{0.98\linewidth}
\centering
  \centerline{\includegraphics[width=0.5\textwidth]{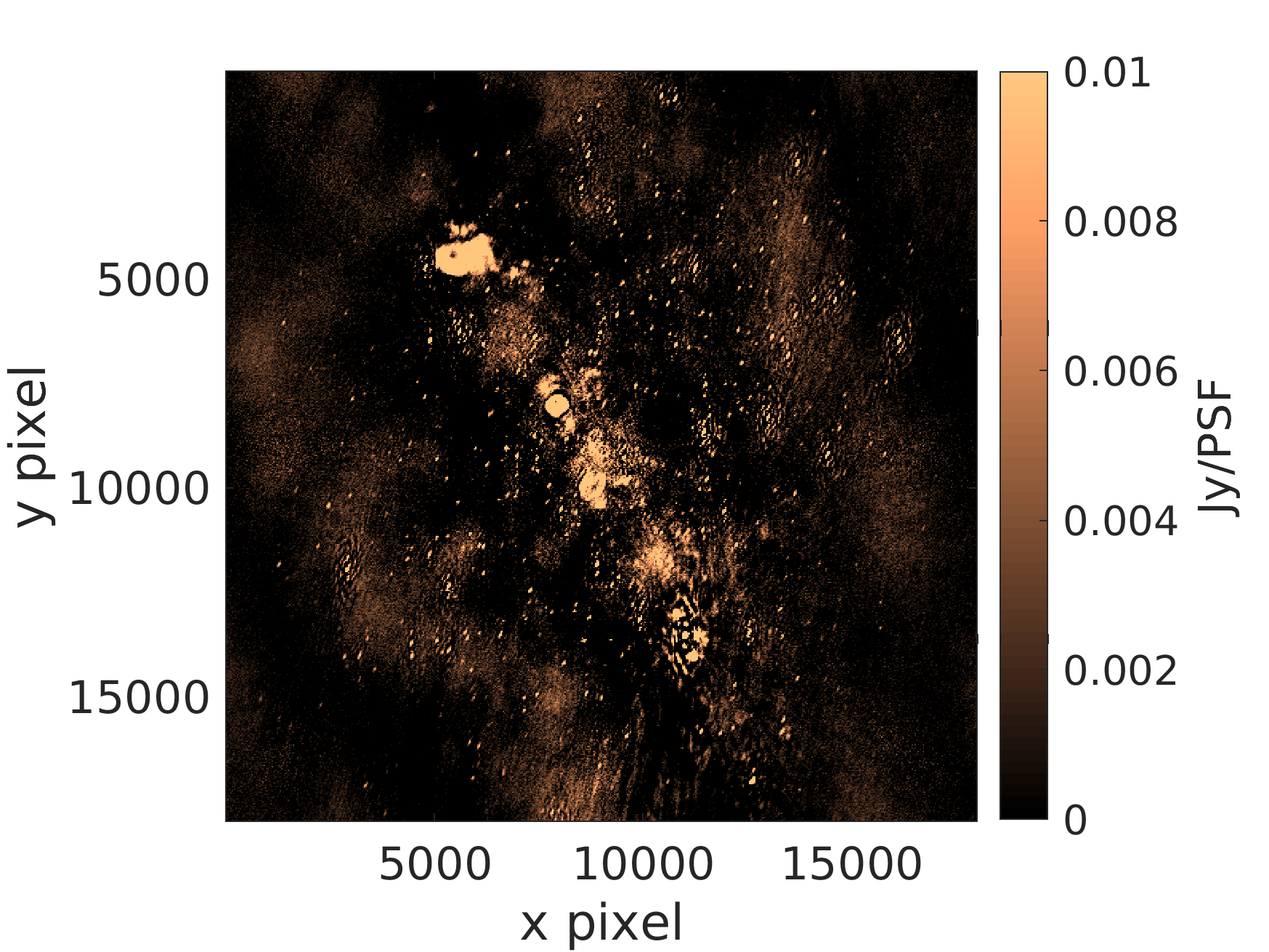}}
\vspace{0.5cm} \centerline{(a)}\smallskip
\end{minipage}\\
\begin{minipage}{0.48\linewidth}
\centering
 \centerline{\includegraphics[width=1.0\textwidth]{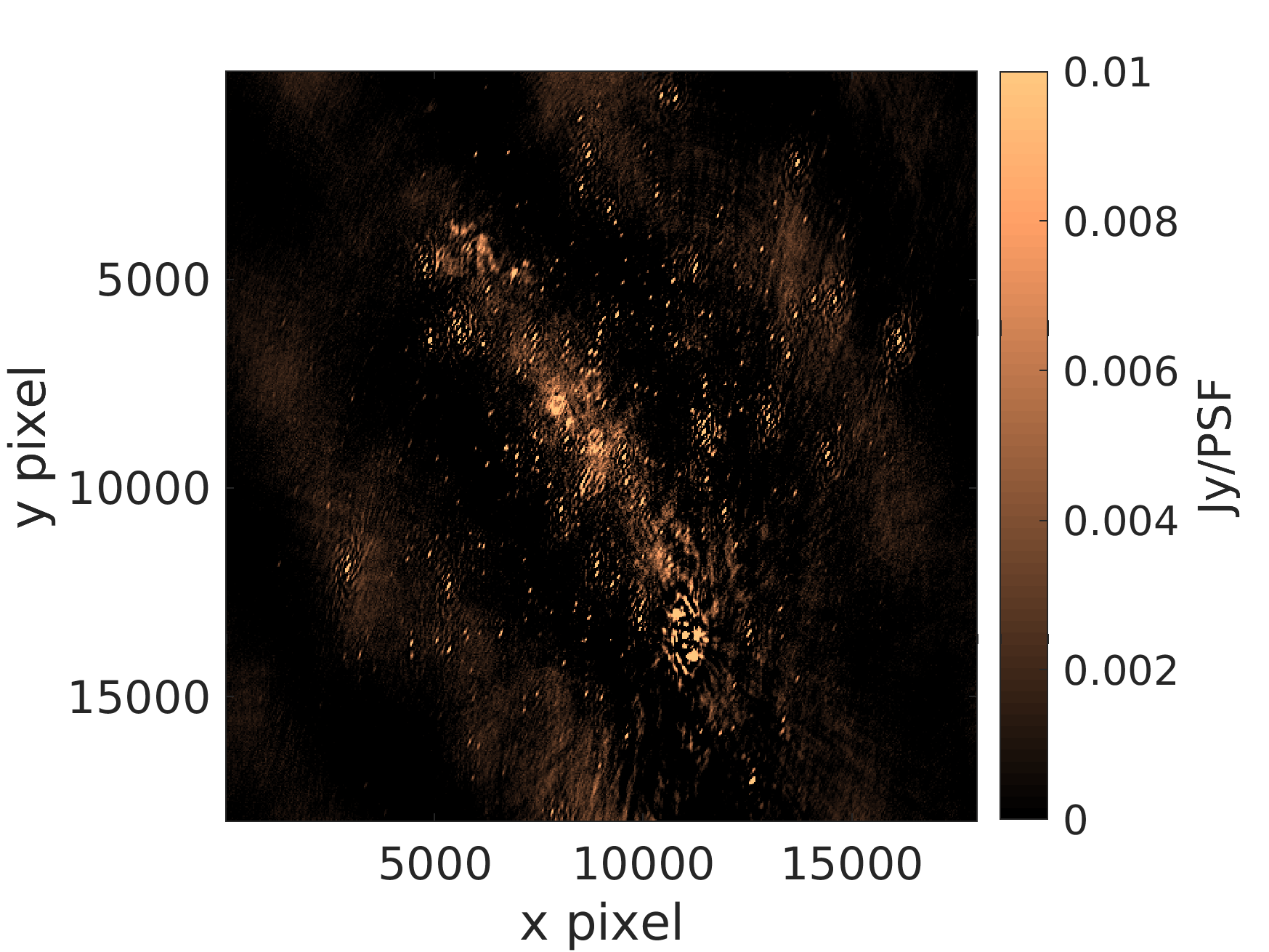}}
\vspace{0.5cm} \centerline{(b)}\smallskip
\end{minipage}
\begin{minipage}{0.48\linewidth}
\centering
 \centerline{\includegraphics[width=1.0\textwidth]{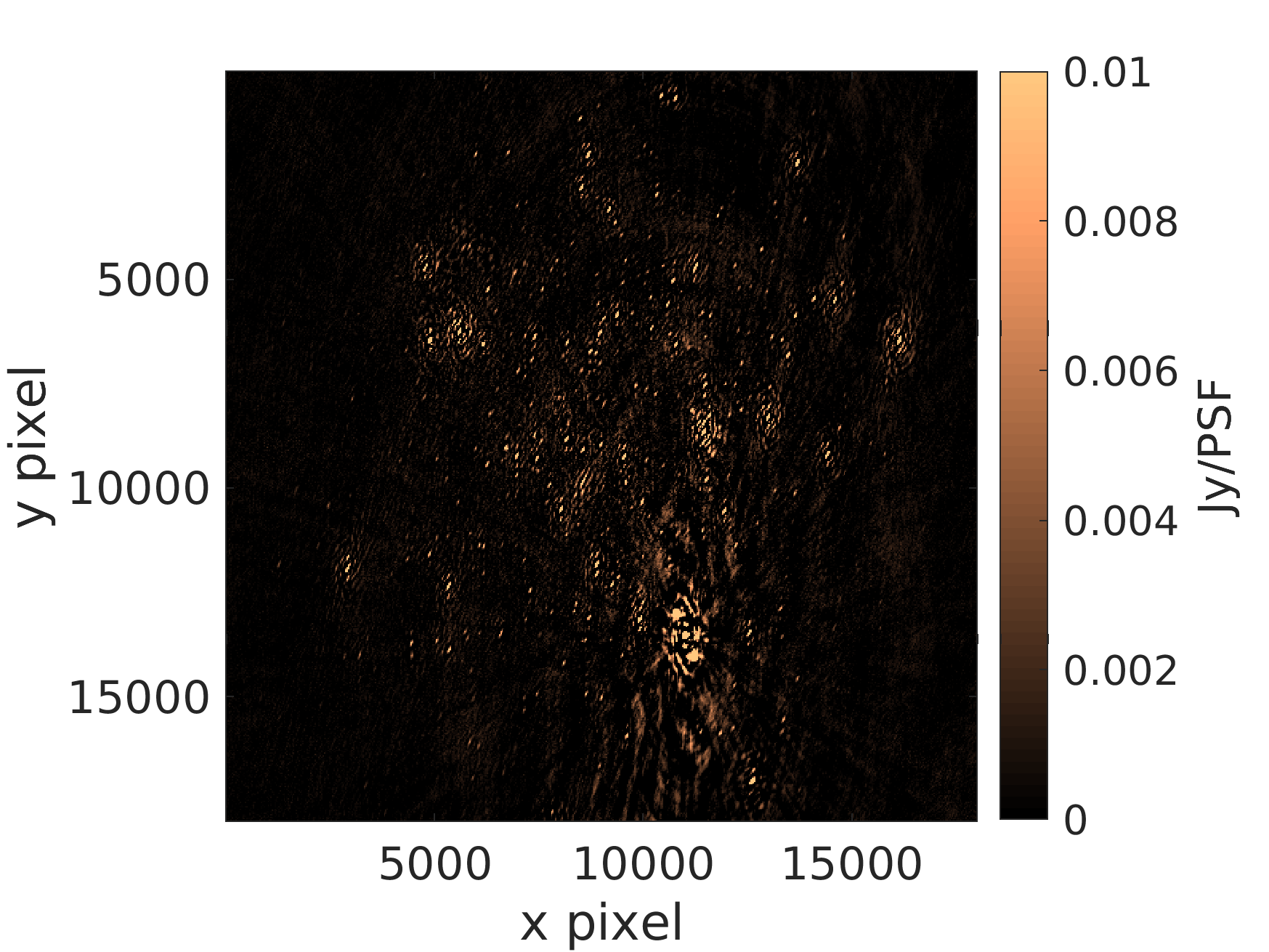}}
\vspace{0.5cm} \centerline{(c)}\smallskip
\end{minipage}\\
\begin{minipage}{0.98\linewidth}
\centering
 \centerline{\includegraphics[width=0.5\textwidth]{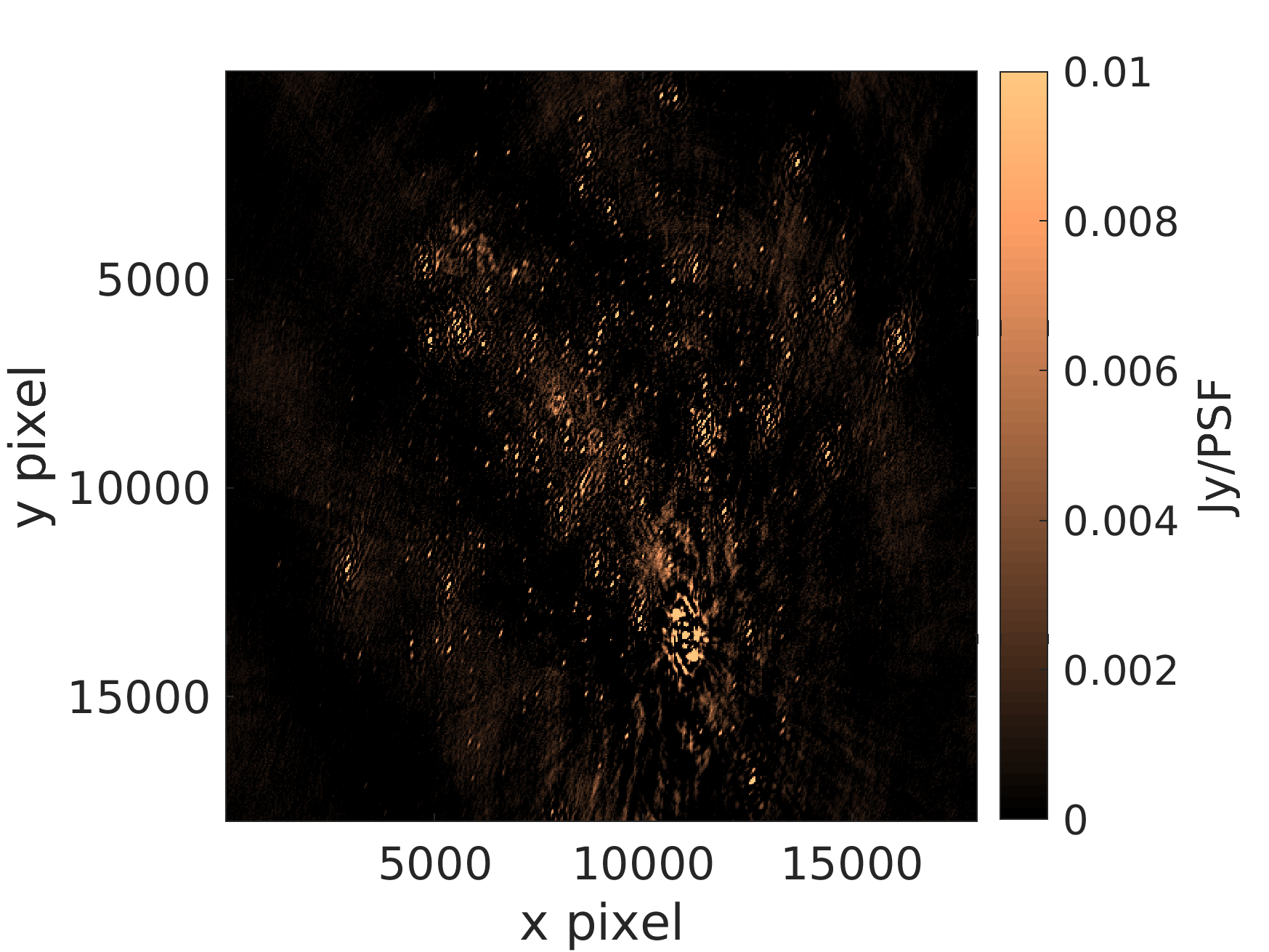}}
\vspace{0.5cm} \centerline{(d)}\smallskip
\end{minipage}
\end{center}
\caption{
  Full field images with $18000 \times 18000$ pixels of size $2^{\prime\prime} \times 2^{\prime\prime}$. The image before calibration is shown in (a). The difference images, i.e., image before calibration $-$ image after calibration, are shown in (b), (c) and (d) for various calibration schemes: (b) calibration with all baselines (c) calibration excluding baselines shorter than 300 wavelengths, and (d) calibration without excluding short baselines and with a model for the large scale diffuse structure. Because we subtract a sky model consisting of only compact sources during calibration, with perfect calibration, the difference images should also show what is being subtracted: the compact sources and their artifacts. However, as we see in (b), we also subtract or suppress a significant fraction of the diffuse structure when we include all baselines in calibration. This suppression is much less in both (c) and (d).
\label{diff_full_field}}
\end{minipage}
\end{figure*}

After calibration, we subtract all compact sources in the sky model to calculate the residual (\ref{residual}) and the residual image obtained for scenario 3 above is shown in Fig. \ref{full_g46} (b). In order to study the performance of the above three scenarios in a quantitative manner, we show zoomed in versions of the residual images in Fig. \ref{zoom_g46} (b),(c),(d) together with the image before direction dependent calibration in Fig. \ref{zoom_g46} (a). Furthermore, in Fig. \ref{diff_full_field} we show the difference images, i.e., the image after calibration subtracted from the image before calibration. 

Visually, images in Fig. \ref{zoom_g46} (b), (c) and (d) look similar, however we are interested in the quantitative differences. In order to do that, we select a small area in these images that have no compact sources but do have diffuse structure (shown by the red square). We evaluate the total flux within this square before and after calibration for all three aforementioned calibration schemes. We see that we recover about 50\% of the flux in calibration scenario 1, where all baselines were included in calibration. Noteworthy is that by excluding baselines shorter than 300 wavelengths, we recover 104\% of the flux, that is an increase which is not physically possible. With the inclusion of a model for the large scale diffuse structure, we recover 88\% of the flux even when we include all baselines in calibration. This is also in agreement with the results obtained using simulated data in section \ref{sec:results} (see Fig. \ref{noise_corr} (a)). Note also that the difference images in Fig. \ref{diff_full_field} qualitatively show the suppression of diffuse structure due to calibration  using all baselines without including a model for the diffuse structure.

\begin{figure*}
\begin{minipage}{0.98\linewidth}
\begin{center}
\begin{minipage}{0.98\linewidth}
\centering
  \centerline{\includegraphics[width=0.5\textwidth]{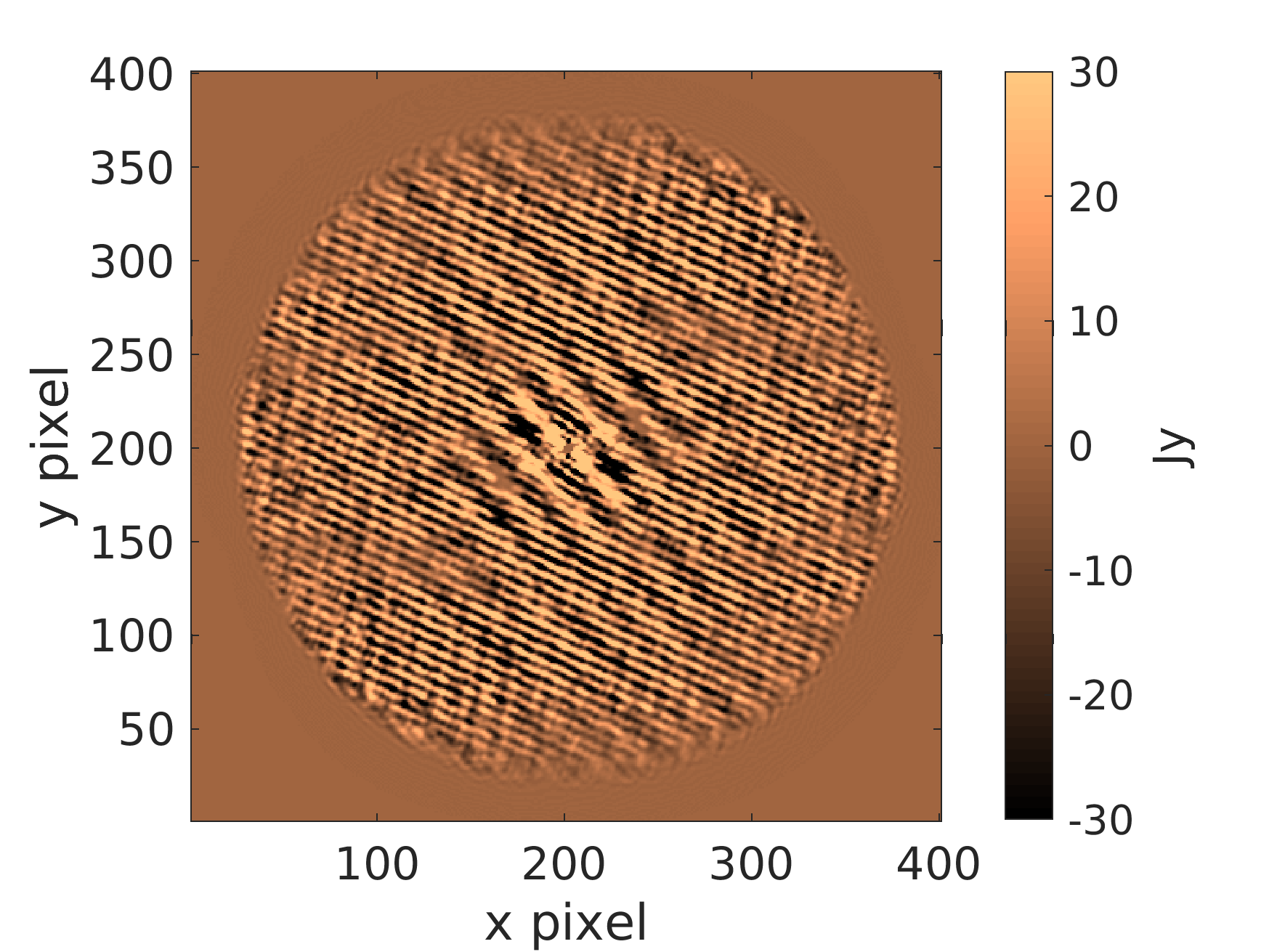}}
\vspace{0.5cm} \centerline{(a)}\smallskip
\end{minipage}\\
\begin{minipage}{0.48\linewidth}
\centering
 \centerline{\includegraphics[width=1.0\textwidth]{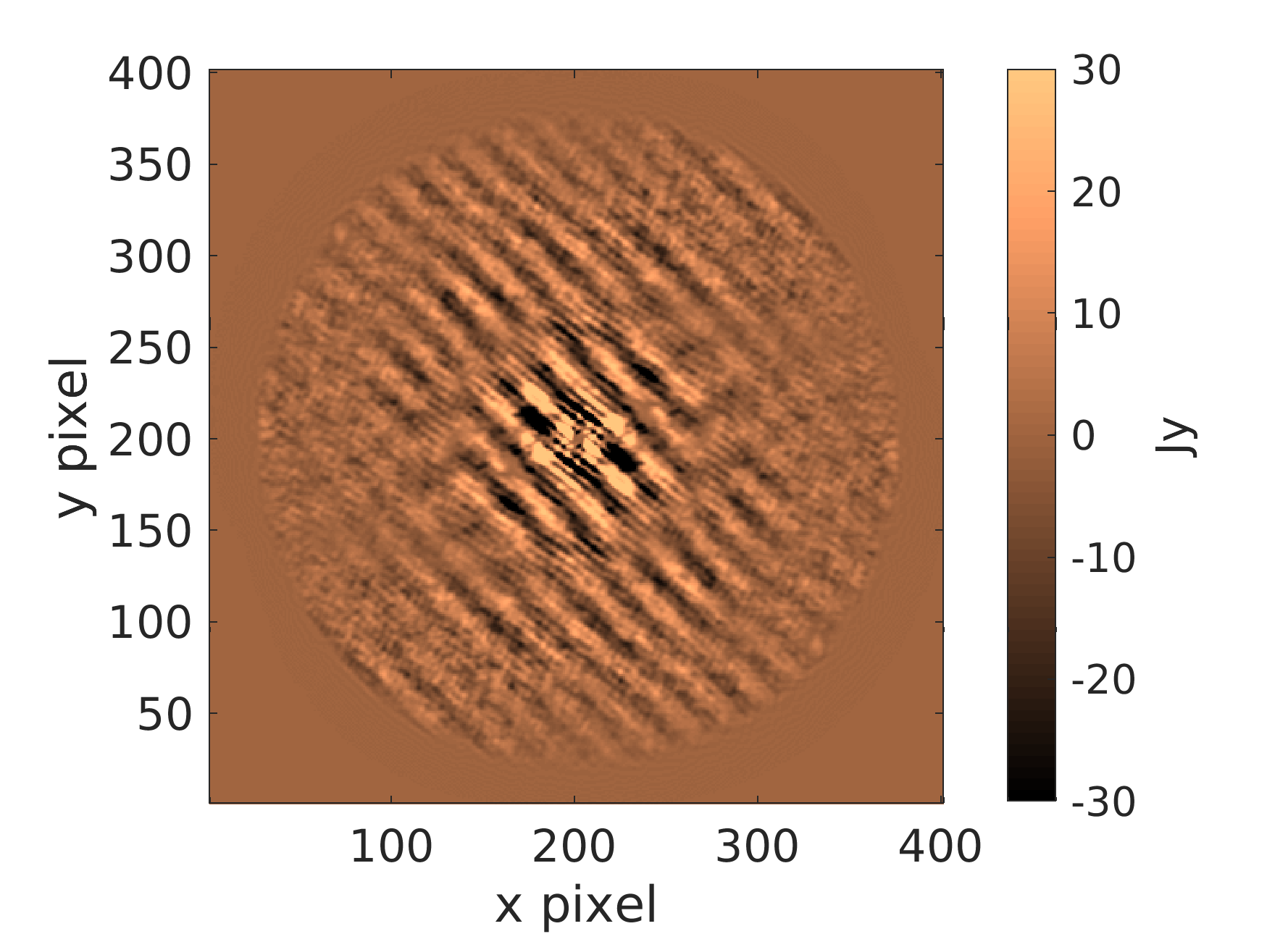}}
\vspace{0.5cm} \centerline{(b)}\smallskip
\end{minipage}
\begin{minipage}{0.48\linewidth}
\centering
 \centerline{\includegraphics[width=1.0\textwidth]{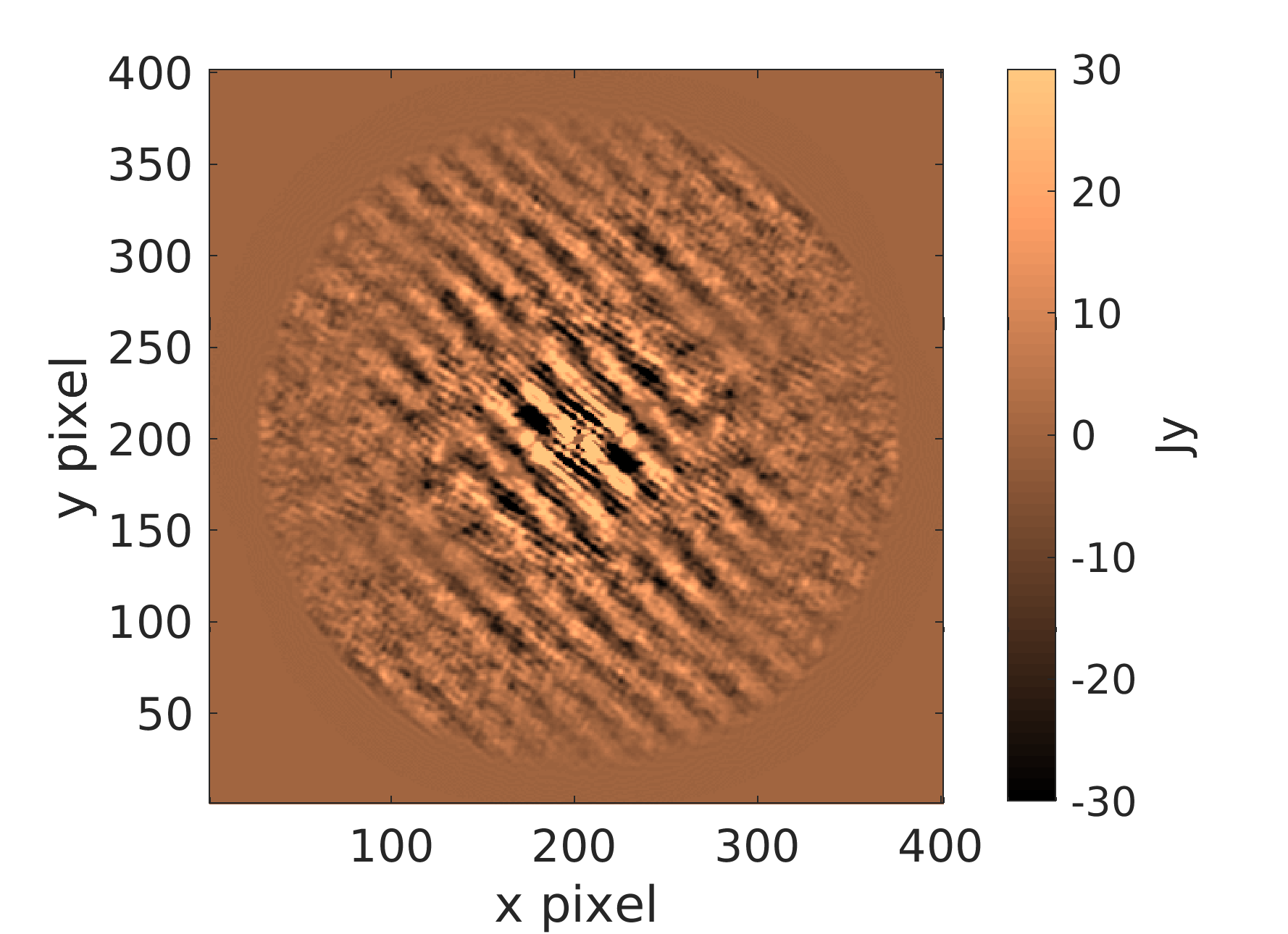}}
\vspace{0.5cm} \centerline{(c)}\smallskip
\end{minipage}\\
\begin{minipage}{0.48\linewidth}
\centering
 \centerline{\includegraphics[width=1.0\textwidth]{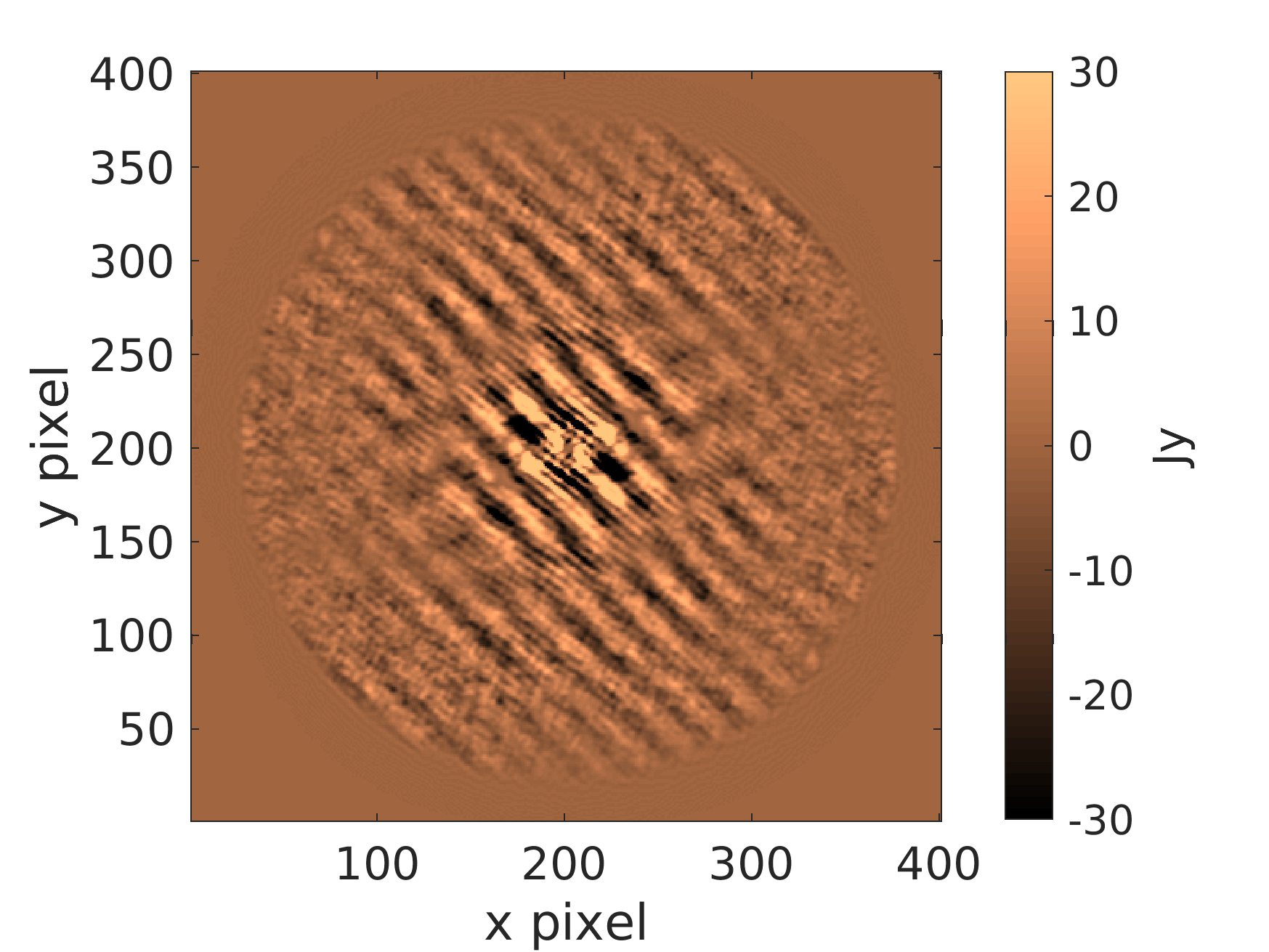}}
\vspace{0.5cm} \centerline{(d)}\smallskip
\end{minipage}
\end{center}
\caption{
  Gridded visibility data (Stokes I, real part) covering a disk of radius 600 wavelengths, (a) before calibration (b) calibration with all baselines (c) calibration excluding baselines shorter than 300 wavelengths, and (d) calibration with all baselines and with the inclusion of a model for the large scale diffuse structure.
\label{gr}}
\end{minipage}
\end{figure*}

The increase in flux recovered when the short baselines were excluded in calibration is something that needs an explanation. In order to do that, we show the gridded visibility data (real part of Stokes I) in Fig. \ref{gr}. Note that we only show the gridded visibilities within a radius of 600 wavelengths in Fig. \ref{gr}. In Fig. \ref{gr} (a), the gridded data before calibration is shown and in Fig. \ref{gr} (b), (c), and (d), the gridded data after calibration is shown for the three calibration scenarios.

\begin{figure*}
\begin{minipage}{0.98\linewidth}
\begin{center}
\begin{minipage}{0.98\linewidth}
\centering
  \centerline{\includegraphics[width=0.5\textwidth]{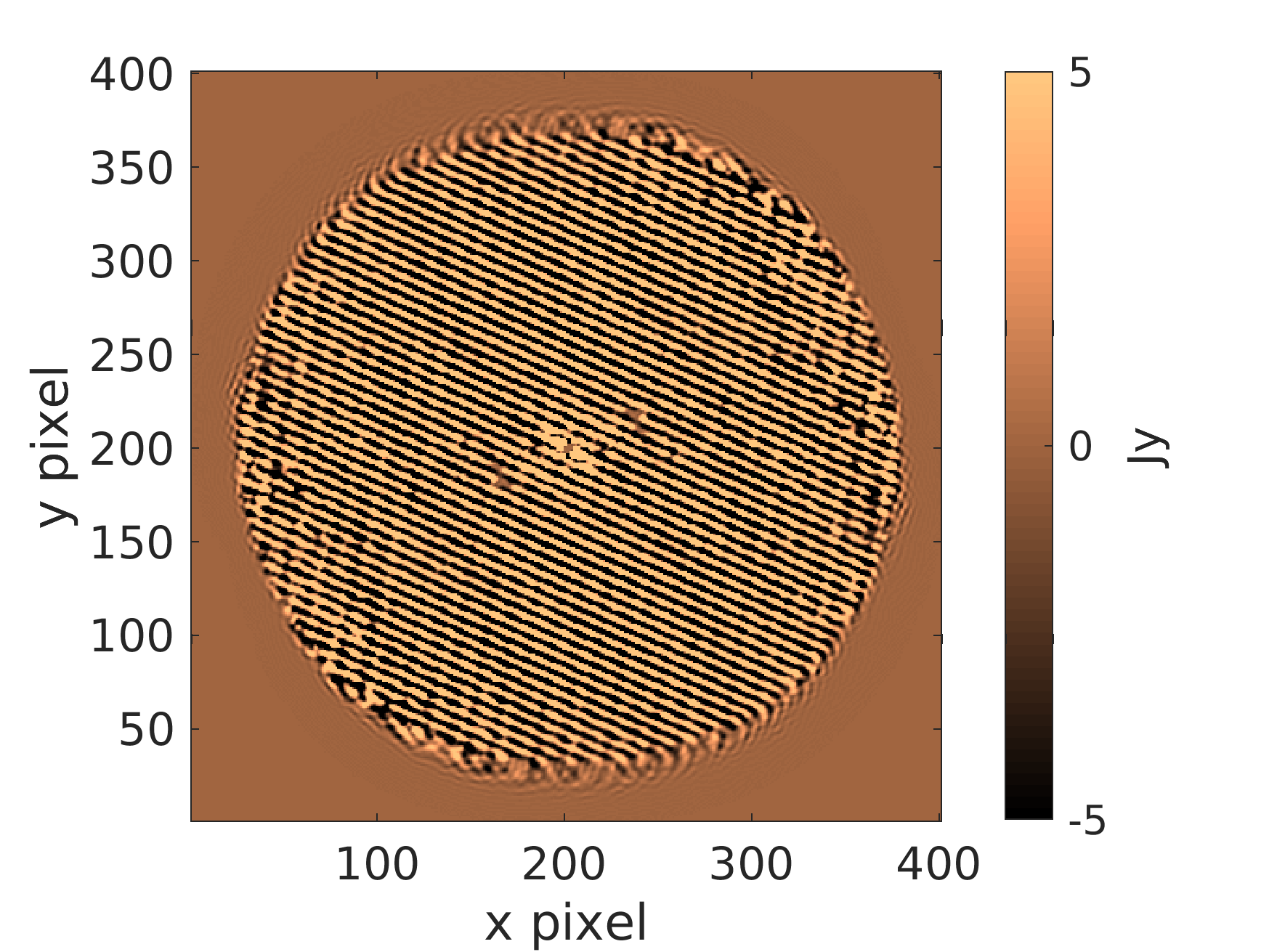}}
\vspace{0.5cm} \centerline{(a)}\smallskip
\end{minipage}\\
\begin{minipage}{0.48\linewidth}
\centering
 \centerline{\includegraphics[width=1.0\textwidth]{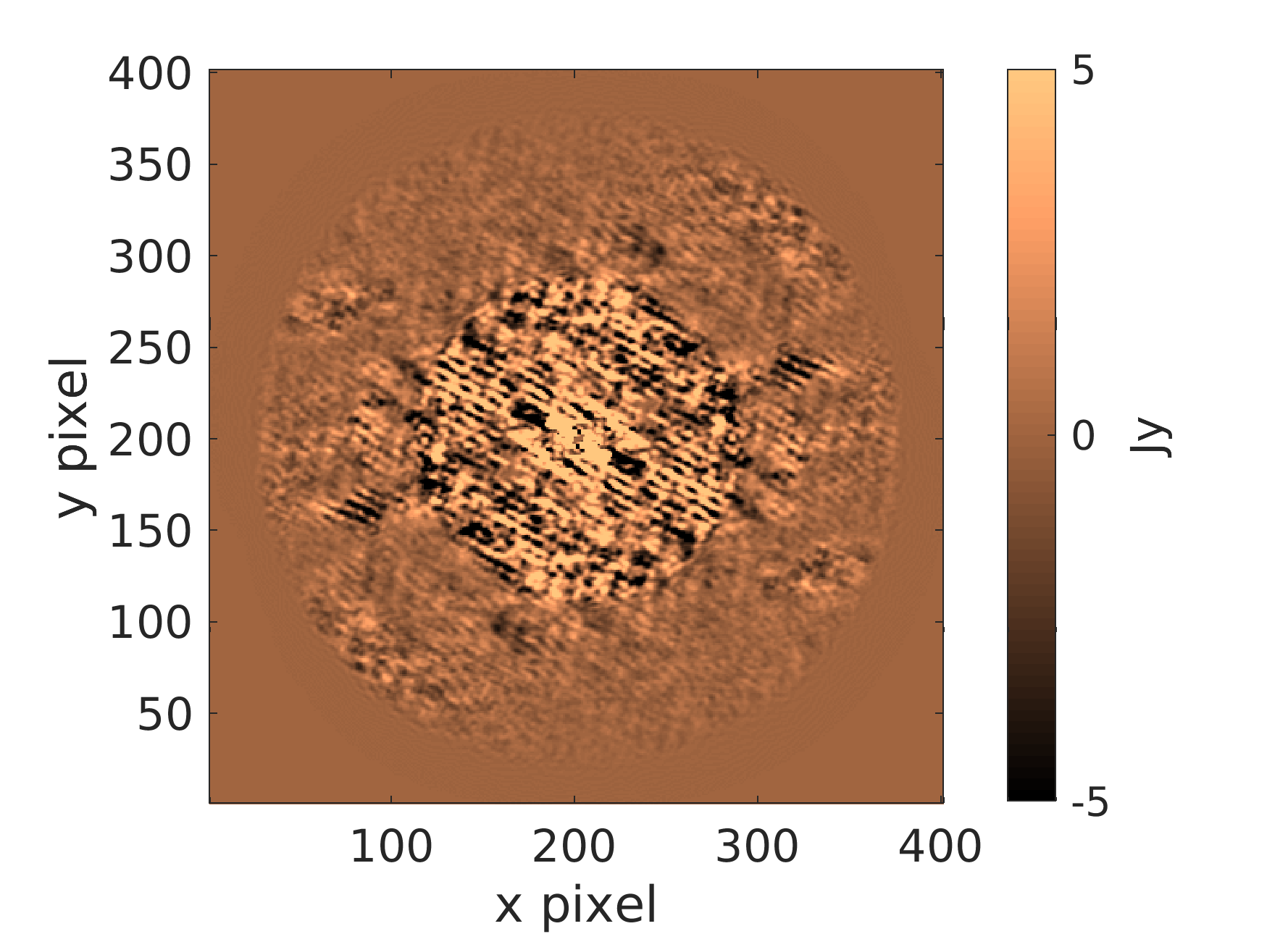}}
\vspace{0.5cm} \centerline{(b)}\smallskip
\end{minipage}
\begin{minipage}{0.48\linewidth}
\centering
 \centerline{\includegraphics[width=1.0\textwidth]{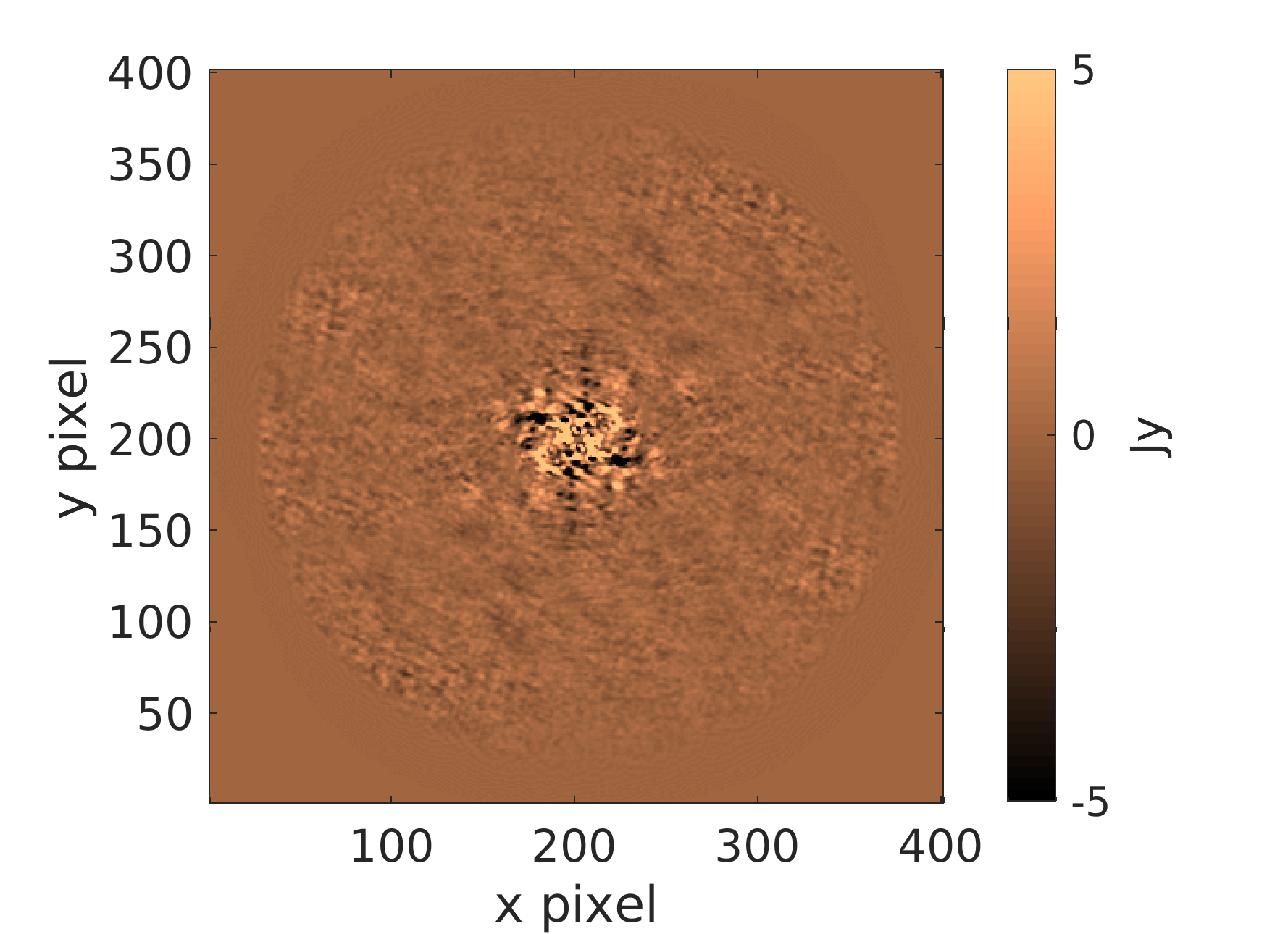}}
\vspace{0.5cm} \centerline{(c)}\smallskip
\end{minipage}
\end{center}
\caption{
  Difference in the gridded visibilities where gridded visibilities of the data after calibration using all baselines were subtracted from gridded visibilities in all other scenarios (a) data before calibration (b) calibration excluding baselines shorter than 300 wavelengths, and (c) calibration including all baselines and with a model for the large scale diffuse structure. An inner disk of radius 300 wavelengths is clearly visible in (b), indicating an artificial increase in the values of this gridded visibilities within this disk.
\label{gr_diff}}
\end{minipage}
\end{figure*}

The differences in Fig. \ref{gr} (b),(c) and (d) are not entirely obvious. In order to highlight the differences if any, we subtract the gridded visibilities after calibration with all baselines (i.e., Fig. \ref{gr} (b)) from all other gridded visibilities as shown in Fig. \ref{gr_diff}. What is seen in Fig. \ref{gr_diff} (a) is primarily the sky model dominated by compact sources that is subtracted to find the residual in (\ref{residual}). A more subtle effect is seen in Fig. \ref{gr_diff} (b), which is the difference between calibration including all baselines and calibration excluding baselines shorter than 300 wavelengths. We clearly see a disk appearing with radius equal to 300 wavelengths in Fig. \ref{gr_diff} (b). This is due to the statistical 'leverage' \citep{Patil2016}. Due to this leverage effect, the residual visibilities within the 300 wavelength radius get artificially boosted, explaining the increase of flux in Fig. \ref{zoom_g46} (c). The inclusion of a diffuse sky model and using all baselines in calibration does not suffer from this effect, as seen in Fig. \ref{gr_diff} (c).

Based on the results in this section we conclude that the inclusion of a model for the large scale diffuse structure during calibration minimizes the suppression of such large scale structure. Excluding short baselines can also minimize this suppression, however it suffers from statistical leverage, artificially boosting the short baseline signals.

\section{Conclusions\label{sec:conclusions}}
We have proposed methods to build and include large scale diffuse sky models in processing of radio interferometric data using shapelet basis functions. Simulations and results based on real data show that this method minimizes the suppression of large scale diffuse structure in the data during direction dependent calibration, without incurring significant computational cost. Therefore, this method will enable the inclusion of short baselines in calibration, increasing the number of usable constraints and improving the end result.

Future work in this topic will focus on adapting the proposed techniques to basis functions other than shapelets as well as using the APC algorithm in other areas in radio interferometric data processing such as in image deconvolution.

The following software already implement the algorithms described in this paper:
\textsc{SAGECal} (\url{https://sagecal.sourceforge.net/}),\\
\textsc{ShapeletGUI} (\url{https://github.com/SarodYatawatta/shapeletGUI}),\\
\textsc{ExCon Imager} (\url{https://sourceforge.net/projects/exconimager/}).

\begin{acknowledgements}
  We thank I. Polderman, M. Haverkorn and Marco Iacobelli for providing us the LOFAR data used in section \ref{sec:observations} and for assisting with accessing the LOFAR archive. We thank the editor and reviewer for valuable comments. LOFAR, the Low Frequency Array designed and constructed by ASTRON, has facilities in several countries, owned by various parties (each with their own funding sources), and collectively operated by the International LOFAR Telescope (ILT) foundation under a joint scientific policy.
\end{acknowledgements}

\bibliographystyle{aa}
\bibliography{references}

\end{document}